\documentclass[a4paper,11pt]{article}
\pdfoutput=1
\usepackage{jheppub}
\usepackage[T1]{fontenc}
\usepackage[english]{babel}
\usepackage{amsthm}
\DeclareMathOperator{\Tr}{Tr}
\DeclareMathOperator{\sign}{sign}
\DeclareMathOperator{\diag}{diag}
\usepackage{tensor}
\usepackage{dsfont}
\usepackage{tikz}
\usetikzlibrary{calc,shapes,arrows}
\newcommand{\HIDDEN}[1]{}
\newcommand{\compliI}{{\scriptscriptstyle\mathrm I}}
\newcommand{\compliR}{{\scriptscriptstyle\mathrm R}}
\usepackage[textsize=small,shadow]{todonotes}
\usepackage{ulem}

\title{\boldmath Consistent Compactification of Double Field Theory\\
on Non-geometric Flux Backgrounds}   
\preprint{LMU-ASC 85/13\\MPP-2013-315}

\author[a]{Falk Hassler}
\emailAdd{f.hassler@lmu.de}
\author[a,b]{and Dieter L\"ust}
\emailAdd{dieter.luest@lmu.de}

\affiliation[a]{Arnold-Sommerfeld-Center f\"ur Theoretische Physik\\
Department f\"ur Physik, Ludwig-Maximilians-Universit\"at M\"unchen\\
Theresienstra\ss e 37, 80333 M\"unchen, Germany}
\affiliation[b]{Max-Planck-Institut f\"ur Physik\\
F\"ohringer Ring 6, 80805 M\"unchen, Germany}

\abstract{In this paper, we construct non-trivial solutions to the $2D$-dimensional field equations of Double Field Theory (DFT) by using a consistent Scherk-Schwarz ansatz. The ansatz identifies $2(D-d)$ internal directions with a twist $U\indices{^M_N}$ which is directly connected to the covariant fluxes $\mathcal{F}\indices{_A_B_C}$. It exhibits $2(D-d)$ linear independent generalized Killing vectors $K\indices{_I^J}$ and gives rise to a gauged supergravity in $d$ dimensions. We analyze the covariant fluxes and the corresponding gauged supergravity with a Minkowski vacuum. We calculate fluctuations around such vacua and show how they gives rise to massive scalars field and vectors field with a non-abelian gauge algebra. Because DFT is a background independent theory, these fields should directly correspond the string excitations in the corresponding background. For $(D-d)=3$ we perform a complete scan of all allowed covariant fluxes and find two different kinds of backgrounds: the single and the double elliptic case. The later is not T-dual to a geometric background and cannot be transformed to a geometric setting by a field redefinition either. While this background fulfills the strong constraint, it is still consistent with the Killing vectors depending on the coordinates and the winding coordinates, thereby giving a non-geometric patching. This background can therefore not be described in Supergravity or Generalized Geometry.}

\begin{document}
\maketitle

\section{Introduction}
String theory has several remarkable features. Most interesting are those that are not present for point particles, but are rather linked to the extended nature of string, like the appearance of stringy symmetries.These are often discovered when compactifying a ten-dimensional superstring theory down to lower dimensions. One prominent example of a stringy symmetry, which becomes manifest during the compactification process, is T-duality. It relies on the existence of string winding modes. By interchanging winding and momentum excitations, T-duality links very small and very large compact dimensions as being completely indistinguishable. Moreover T-duality allows for the existence of new `geometries' as consistent string backgrounds. These are certain generalizations of standard Riemannian spaces and often called non-geometric string backgrounds \cite{Dabholkar:2005ve}. The dynamics of a string in such a non-geometric background is governed by the interplay between winding and momentum modes. This gives rise to many new phenomena which are not present in a geometric background with momentum modes only. One prominent example for such new effects is a new kind of spatial non-commutativity and non-associativity of the form $[ X^I(\tau, \sigma), X^J(\tau, \sigma)]\simeq P^K$ resp. $[ [X^I(\tau, \sigma), X^J(\tau, \sigma),X^K(\tau, \sigma)]]\neq 0$ of the closed string coordinates in the presence of non-geometric $Q$- and $R$-fluxes, as has been argued in \cite{Blumenhagen:2010hj,Lust:2010iy,Blumenhagen:2011ph,Condeescu:2012sp,Andriot:2012vb,Mylonas:2012pg,Bakas:2013jwa,Blumenhagen:2013zpa,Mylonas:2013jha}. In correspondence to Heisenberg's well know uncertainty relation between position and momentum, these relations describe a stringy limited resolution of the string's position, which can be interpreted as a fuzzy non-commutative and non-associative space. These effects arise on the interface between large and small compact dimensions, which are very different for a string compared to a point particle. Furthermore non-geometric backgrounds extend the landscape of string theory considerably and perhaps help to find one day a string compactification which reproduces the phenomenology of our universe. Thus it is important to understand the properties of such backgrounds in more detail.

In this paper we want to discuss the construction of non-geometric backgrounds and analyze their spectrum in type IIA/IIB superstring theory. We focus on the NS/NS sector, which consists of three different massless string excitations: the symmetric metric $g_{ij}$, the antisymmetric $B$-field $B_{ij}$ and a scalar $\phi$ called dilaton. Their complete dynamics are governed by string field theory in $D$-dimensions. But in general, string field theory is much too involved to be evaluated explicitly. Hence an effective field theory is used in the low energy limit. It is defined by the following action
\begin{equation}\label{eqn:nsnsaction}
  S_\mathrm{NS} = \int\mathrm{d}^{D}x\, \sqrt{-g} e^{-2\phi} \left(\mathcal R +
    4 \partial_\mu \phi \partial^\mu \phi - \frac{1}{12} H_{\mu\nu\rho} H^{\mu\nu\rho} \right)\,.
\end{equation}
which describes the NS/NS sector of a $\mathcal{N}=2$ supergravity. Due to its construction, this effective theory only considers strings with momentum modes. In order not to violate the low energy limit, the compact dimensions described by \eqref{eqn:nsnsaction} have to be large. Due to this limitation the stringy symmetries, in particular T-duality, are not implemented into this action. Because non-geometric backgrounds depend on the interplay between winding and momentum modes, this action is only of limited use when studying the properties of non-geometric backgrounds. Thus the fields $g_{ij}$, $B_{ij}$ and $\phi$ are in generally ill defined (either globally or even locally) for a non-geometric background. For non-geometric backgrounds which are T-dual to geometric ones, a fields redefinition can be performed to obtain a well defined geometric description \cite{Andriot:2011uh,Andriot:2012wx,Andriot:2012an,Blumenhagen:2012nk,Blumenhagen:2012nt,Blumenhagen:2013aia,Andriot:2013xca}. But for all other non-geometric backgrounds, which are in the following called truly non-geometric, this is not possible. 

Double Field Theory (DFT) \cite{Siegel:1993th,Hull:2009mi,Hull:2009zb,Hohm:2010jy,Hohm:2010pp} is a promising approach to overcome these problems. In particular DFT allows us to make T-duality a manifest symmetry of the effective theory. Hence, we will investigate consistent Scherk-Schwarz like dimensional reductions \cite{Scherk:1978ta,Scherk:1979zr} of the $2 D$-dimensional DFT \cite{Aldazabal:2011nj,Grana:2012rr}. Recently, such reductions were also discussed in the context of generalized geometry \cite{Lee:2014mla}. They give rise to a gauged supergravity in the remaining $d$-dimension, exhibiting non-Abelian gauge symmetries together with a scalar potential on their moduli space (parameters which describe the shape of the background in the internal direction). This potential can be used to stabilize some of the moduli and so remove a lot of arbitrariness when choosing the explicit shape of a background. Furthermore the scalar potential possesses phenomenologically interesting properties, like a non-vanishing cosmological constant \cite{Samtleben:2008pe}. Similar effects arise in massive type II theories, which were discussed in DFT \cite{Hohm:2011cp}, too. We find solutions for the field equations of the $d$-dimensional gauged supergravity and lift them up to solution of the full DFT field equations $\mathcal{R}_{MN}=0$. Here $\mathcal{R}_{MN}$ is the generalized Ricci tensor of the double geometry, in $2 D$ dimensions. This uplift is possible when the Scherk-Schwarz ansatz exhibits $2 (D-d)$ Killing vectors \cite{Scherk:1978ta,Scherk:1979zr,Pons:2003ka}.

Among all the different gauged supergravities, that can arise for a Scherk-Schwarz ansatz, we are focusing on the ones with a Minkowski vacuum. Such theories exhibit a minimum of the scalar potential, on which the scalar potential vanishes. This restriction puts additional constraints on the covariant fluxes $\mathcal{F}_{ABC}$  \cite{Aldazabal:2011nj,Geissbuhler:2013uka}, which specifies the explicit form of the Scherk-Schwarz ansatz. As we will show, these fluxes are directly connected to fluxes $H_{ijk}$, $f^i_{jk}$, $Q^{ij}_k$ and $R^{ijk}$ which are widely used to characterize non-geometric backgrounds. Similar calculations were discussed e.g. in \cite{Geissbuhler:2013uka,Blumenhagen:2013hva}. For gauged supergravities with Minkowski vacuum, we discuss small fluctuations around the vacuum. This gives rise to $(D-d)^2$ scalar field and $2(D-d)$ vector gauge bosons. We calculate the masses of the scalars and the gauge group of the vectors. Because DFT is constructed as a background independent low energy description of string theory, the spectrum we have obtained in this way should be identical to CFT calculations, but we leave an explicit verification to future work.

In order to provide explicit examples for non-geometric spaces, we restrict ourself to $(D-d)=3$ internal dimensions. Here we provide all supergravities with Minkowski vacuum and consistent uplift. There are only two of them, which we call single elliptic and double elliptic case. The double geometries in the internal direction of both cases correspond to fibrations, where the doubled fiber is a four-dimensional torus $\mathrm{T}^4$ over a doubled circle as base. The double elliptic case is not T-dual to a geometric description, an its generalized geometric description within DFT has been discussed in \cite{ReviewDFT:2013hlb}. It exhibits $H$-, $f$- and $Q$-flux at the same time. Nevertheless it is compatible with the strong constraint of DFT. Thus it is a truly non-geometric space. It cannot be written in terms of a globally well defined metric, $B$-field and dilaton. Nevertheless, as discussed recently, this notion of non-geometric backgrounds can be properly defined in DFT \cite{Andriot:2012an}. In particular generalized coordinate transformations can be used as the so-called patching conditions for non-geometric spaces. This is of particular importance for truly non-geometric spaces that are not T-dual to any geometric spaces. In fact, without the use of the  DFT formalism,  the dimensional reduction on these non-geometric backgrounds could not have been discussed so far, a fact, which clearly demonstrates the necessity to go beyond the standard effective string action, when one wants to explore the full landscape \cite{Susskind:2003kw,Ashok:2003gk} of consistent string compactifications. We give explicit expressions for the Killing vectors, the twist of the Scherk-Schwarz ansatz, the masses of scalar bosons and the structure coefficients of the gauge boson's gauge group. All these results are in accordance with the CFT calculation for a asymmetric orbifold presented in \cite{Condeescu:2013yma}. Thus we conjecture that the double elliptic case is the low energy description of superstring theory in this background. This shows, that non-geometric background are not a mere theoretical construct, but leads to effective theories which are beyond the case of SUGRA.

The paper is organized as follows: In section~\ref{sec:DFTreview}, we review some important features and notions of DFT, needed throughout the paper. Section~\ref{sec:DFTbackgrounds} defines the Scherk-Schwarz ansatz in terms of the twist $U\indices{^M_N}$ and connects this twist to the covariant fluxes $\mathcal{F}\indices{_A_B_C}$. It discusses several constraints that the covariant fluxes have to fulfill and finally presents the action of the gauged supergravity obtained by the Scherk-Schwarz ansatz. Gauged supergavities with a Minkowski vacuum are discussed in section~\ref{sec:minkowski}. Here further constraints on the covariant fluxes are defined. The masses of the scalar bosons, which arise through fluctuations around the vacuum, are calculated. For $(D-d)=3$ all flux constraints are solved explicitly. Finally, section~\ref{sec:constrtwists} presents the explicit construction of the twist $U\indices{^M_N}$ and the Killing vectors $K\indices{_I^J}$. It also discusses how different values for the $B$-field, the $\beta$-field and the metric arise in the elliptic and double elliptic case through field redefinition. A conclusion about the results in the paper is drawn in section~\ref{sec:conclusions}.

\section{Double field theory}\label{sec:DFTreview}
In this section we review some important properties of DFT, which will be relevant for the calculations in this paper. We start with introducing the DFT action and show its various symmetries. Afterwards we present the equations of motion which arise from the variation of this action. Finally we discuss how fluxes arise in DFT.

\subsection{Action and its symmetries}\label{sec:dftandsym}
DFT is an effective description of closed string theory that takes into account both momentum and winding modes in compact space time. Hence in addition to the $D$ space time coordinates $x$ (conjugate to the momentum modes), it introduces $D$ new coordinates $\tilde x$ (conjugate to the winding modes of the string). In total there are now $2 D$ coordinates which are combined into the $2D$-dimensional vector $X^M=\begin{pmatrix}\tilde x_i & x^i\end{pmatrix}$. To lower and raise the index $M$ of this vector, the O$(D,D)$ invariant metric
\begin{equation}\label{eqn:etaMN}
    \eta_{MN}=\begin{pmatrix}
      0 & \delta^i_j \\
      \delta_i^j & 0
    \end{pmatrix} \quad \text{and its inverse} \quad
    \eta^{MN}=\begin{pmatrix}
      0 & \delta_i^j \\
      \delta^i_j & 0
    \end{pmatrix}
\end{equation}
are used. Furthermore one defines the partial derivative according to $\partial^M=\begin{pmatrix}\partial_i & \tilde{\partial}^i\end{pmatrix}$. Now the DFT action can be expressed in the generalized metric formulation \cite{Hohm:2010pp} as
\begin{equation}\label{eqn:dftaction}
  S_\mathrm{DFT} = \int \mathrm{d}^{2D} X\,e^{-2\phi'} \mathcal{R}
\end{equation}
where
\begin{align}\label{eqn:genricciscalar}
  \mathcal{R} = 4 \mathcal{H}^{MN} \partial_M \phi' \partial_N \phi'
    - \partial_M \partial_N \mathcal{H}^{MN} &- 4\mathcal{H}^{MN}
      \partial_M \phi' \partial_N \phi' + 4 \partial_M \mathcal{H}^{MN}
      \partial_N \phi' \nonumber \\
  + \frac{1}{8} \mathcal{H}^{MN} \partial_M \mathcal{H}^{KL}  
    \partial_N \mathcal{H}_{KL} &- \frac{1}{2} \mathcal{H}^{MN}\partial_N
    \mathcal{H}^{KL}\partial_L\mathcal{H}_{MK}
\end{align}
is called the generalized Ricci or curvature scaler and
\begin{equation}\label{eqn:genmetricBg}
  \mathcal{H}^{MN}=\begin{pmatrix}
    g_{ij} - B_{ik}g^{kl}B_{lj} & -B_{ik}g^{kj} \\
    g^{ik} B_{kj} & g^{ij}
  \end{pmatrix}
\end{equation}
is the generalized metric. It combines the metric $g_{ij}$ and the $B$-field $B_{ij}$ into a O$(D,D)$ valued, symmetric tensor with the properties
\begin{equation}
  \mathcal{H}^{MN} \eta_{ML} \mathcal{H}^{LK} = \eta^{NK} \quad \text{and}
  \quad \mathcal{H}^{MN} = \mathcal{H}^{NM} \,.
\end{equation}
The dilaton $\phi$ is encoded in the O$(D,D)$ singlet
\begin{equation}
  \phi' = \phi - \frac{1}{2} \log \sqrt{-g}\,.
\end{equation}
Because it only consists of covariant quantities, the action \eqref{eqn:dftaction} posses a manifest, global $O(D,D)$ symmetry. The symmetry is global only, but the DFT action \eqref{eqn:dftaction} has further symmetries which are local.

In order to display one of them, we express the generalized metric in terms of the generalized vielbein $E\indices{^A_M}$, employing a vielbein formalism, as originally introduced by Siegel in \cite{Siegel:1993th}  and applied to DFT in \cite{Hohm:2010xe}. We thus express the generalized metric in terms of frame fields via 
\begin{equation}\label{eqn:genmetricformvielbein}
  \mathcal{H}^{MN} = E\indices{^A_M} \delta_{AB} E\indices{^B_N}\, . 
\end{equation}
In the following it is convenient to slightly adapt the frame formalism of \cite{Siegel:1993th,Hohm:2010xe} in such a way that the frame field can be viewed as a proper group element, as has been used in \cite{Geissbuhler:2011mx}. The flat generalized metric is then given by
\begin{equation}
  \delta_{AB} =
    \begin{pmatrix} \eta^{a b} & 0 \\ 
      0 & \eta_{a b}
    \end{pmatrix}\,,
\end{equation}
where $\eta_{ab}$ and its inverse $\eta^{ab}$ are the usual $D$-dimensional Minkowski metric. From now on we distinguish between the indices $A, B, \dots$ and $M, N, \dots\;$. The former are called flat and the latter curved. As already mentioned, the generalized metric $\mathcal{H}^{MN}$ is an O$(D,D)$ valued tensor, and here the generalized vielbein is O$(D,D)$ valued, too:
\begin{equation}\label{eqn:vielbeinodd}
  E\indices{^A_M} \eta_{MN} E\indices{^B_N} = \eta^{AB}
    \quad \text{with} \quad
  \eta^{AB}=\begin{pmatrix}
    0 & \delta_a^b \\
    \delta^a_b & 0
  \end{pmatrix}\,.
\end{equation}
Here $\eta^{AB}$ in flat indices does not differ for $\eta^{NM}$ in curved ones. Let us now inspect the local Lorentz group in some detail. Consider the local double Lorentz transformation of the generalized vielbein
\begin{equation}\label{eqn:localodxodsym}
  \tilde E\indices{^A_M} = T\indices{^A_B} E\indices{^B_M}\,.
\end{equation}
Requiring that this leaves the generalized metric invariant, the transformation has to fulfill
\begin{equation}\label{eqn:trafoo2d}
  T\indices{^A_C} \delta^{CD} T\indices{^B_D} = \delta^{AB}\,.
\end{equation}
In addition, the transformed generalized vielbein $\tilde E\indices{^A_M}$ has still to satisfy \eqref{eqn:vielbeinodd}, which gives rise to the further constraint
\begin{equation}\label{eqn:trafoodd}
  T\indices{^A_C} \eta^{CD} T\indices{^B_D} = \eta^{AB}\,.
\end{equation}
Transformations that simultaneously solve \eqref{eqn:trafoo2d} and \eqref{eqn:trafoodd}, belong to the local subgroup O$({D-1},1)_\mathrm{R}\times$O$(1,{D-1})_\mathrm{L}$. In order to examine their explicit form, we transform $\eta_{AB}$ into the diagonal form
\begin{gather}\label{eqn:trafobaredind}
  R\indices{^{\bar A}_C} \eta^{CD} R\indices{^{\bar B}_D} =
    \eta^{\bar A\bar B} = \begin{pmatrix}
    - \eta_{\bar a \bar b} & 0 \\
    0 & \eta^{\bar a \bar b} 
  \end{pmatrix} \\
  \text{with} \quad
  R\indices{^{\bar A}_B} = \frac{1}{\sqrt{2}}
  \begin{pmatrix}
    \delta_{\bar a}^b & -\eta_{\bar a b} \\
    \eta^{\bar a b} & \delta^{\bar a}_b 
  \end{pmatrix} \quad \text{and} \quad
  R\indices{_{\bar A}^B} = \frac{1}{\sqrt{2}}
  \begin{pmatrix}
    \delta^{\bar a}_b & -\eta^{\bar a b} \\
    \eta_{\bar a b} & \delta_{\bar a}^b 
  \end{pmatrix}\,.
\end{gather}
Here, bared indices are used in order to distinguish between the different representations of the invariant metric\footnote{It is important to distinguish its notation form the one introduces in \cite{Hohm:2010xe}. In \cite{Hohm:2010xe}, a tensor $T\indices{_{\bar a}^{\bar b}}$ is relates to $T\indices{^{\bar a}_{\bar b}}$ by rising and lowering the bared indices with the Minkowski metric $\eta^{ab}$ and $\eta_{ab}$, respectively. While in our notation, $T\indices{_{\bar a}^{\bar b}}$ and $T\indices{^{\bar a}_{\bar b}}$ are totally unrelated objects.}
. In the same fashion, the bared version
\begin{equation}
  R\indices{^{\bar A}_C} \delta^{CD} R\indices{^{\bar B}_D} =
    \delta^{\bar A\bar B} = \begin{pmatrix}
    \eta_{\bar a \bar b} & 0 \\
    0 & \eta^{\bar a \bar b}
  \end{pmatrix}
\end{equation}
of the flat generalized metric is calculated. The deeper meaning of the coordinate transformation mediated by $R\indices{^{\bar A}_B}$ becomes clear, when one applies it on the doubled coordinates $X^M$ and obtains
\begin{equation}
  R\indices{^{\bar M}_N} X^N = \frac{1}{\sqrt{2}} \begin{pmatrix}
    \tilde x_{\bar i} - x_{\bar i} &
    \tilde x^{\bar i} + x^{\bar i}
  \end{pmatrix} = \begin{pmatrix}
    {x_{\scriptscriptstyle\mathrm R}}_{\bar i} &
    {x_{\scriptscriptstyle\mathrm L}}^{\bar i}
  \end{pmatrix}\,.
\end{equation}
Here $x_{\scriptscriptstyle\mathrm R}$ and $x_{\scriptscriptstyle\mathrm L}$ are the positions conjugated to the momenta of the closed string's right and left moving part. Expressing \eqref{eqn:trafoo2d} and \eqref{eqn:trafoodd} in bared indices gives rise to
\begin{equation}
  \begin{pmatrix}
    T\indices{_{\bar a}^{\bar c}} & T_{\bar a\bar c} \\
    T^{\bar a\bar c} & T\indices{^{\bar a}_{\bar c}} 
  \end{pmatrix}
  \begin{pmatrix}
    \pm \eta_{\bar c\bar d} & 0 \\
    0 & \eta^{\bar c\bar d} \\
  \end{pmatrix}
  \begin{pmatrix}
    T\indices{^{\bar d}_{\bar c}} & T^{\bar d\bar c} \\
    T_{\bar d\bar c} & T\indices{_{\bar d}^{\bar b}} 
  \end{pmatrix}
  =
  \begin{pmatrix}
    \pm \eta_{\bar a\bar b} & 0 \\
    0 & \eta^{\bar a\bar b} \\
  \end{pmatrix}
\end{equation}
which is solved by $T_{\bar a\bar b}=T^{\bar a\bar b}=0$ and two different 
$O(1,D-1)$ transformations 
\begin{equation}
  u\indices{_{\bar a}^{\bar c}} \eta_{\bar c\bar d}
    u\indices{_{\bar b}^{\bar d}} = \eta_{\bar a\bar b}
  \quad \text{and} \quad
  v\indices{^{\bar a}_{\bar c}} \eta^{\bar c\bar d}
    v\indices{^{\bar b}_{\bar d}} = \eta^{\bar a\bar b}\,.
\end{equation}
They are identified with the remaining components $T\indices{^A_B}$ as $T\indices{_{\bar a}^{\bar b}}=u\indices{_{\bar a}^{\bar b}}$ and $T\indices{^{\bar a}_{\bar b}}=v\indices{^{\bar a}_{\bar b}}$. In unbared indices this transformation reads
\begin{equation}\label{eqn:trafoonxon}
  T\indices{^A_B} = 
    R\indices{^A_{\bar C}} T\indices{^{\bar C}_{\bar D}} 
    R\indices{^{\bar D}_B} = 
  \begin{pmatrix}
    u\indices{_a^b} + v\indices{_a^b} &
    u\indices{_a_b} - v\indices{_a_b} \\
    u\indices{^a^b} - v\indices{^a^b} &
    u\indices{^a_b} + v\indices{^a_b}
  \end{pmatrix}\,.
\end{equation}
Hence the generalized metric and therewith the DFT action \eqref{eqn:dftaction} are invariant under local double Lorentz transformations of the form \eqref{eqn:localodxodsym}.

Except for the dilaton, the generalized vielbein combines all fields of the theory. As an element of O$(D,D)$ it has $D(2D-1)$ independent degrees of freedom. By gauging the local double Lorentz symmetry only $D^2$ of them remain. A possible parameterization of the generalized vielbein is given by
\begin{equation}\label{eqn:EAMfixed}
  E\indices{^A_M} = \begin{pmatrix}
    e\indices{_a^i} & e\indices{_a^l} B_{li} \\
    0 & e\indices{^a_i}
  \end{pmatrix}
\end{equation}
in terms of the metric's vielbein $e\indices{^a_i}$ with $e\indices{^a_i} \eta_{ab} e\indices{^b_j} = g_{ij}$ and the antisymmetric $B$-field $B_{ij}$. If $e\indices{^a_i}$ is restricted to be an upper triangular matrix, this parameterization fixes the double Lorentz symmetry completely. An O$(D,D)$ vielbein without any gauge fixing is
\begin{equation}\label{eqn:EAMgeneral}
  E\indices{^A_M} = \begin{pmatrix}
    e\indices{_a^i} & e\indices{_a^l} B_{li} \\
    e\indices{^a_l} \beta^{li} & e\indices{^a_i} +
      e\indices{^a_l} \beta^{lk} B_{ki}
  \end{pmatrix}
\end{equation}
where $e\indices{^a_i}$ is an unrestricted vielbein of $g_{ij}$ and $\beta^{ij}$ is an antisymmetric bi-vector.

Finally, the DFT action is also invariant under generalized diffeomorphisms. These transform $X^M$ into $\tilde{X}^M=X^M - \xi^M$ where $\xi^M$ is infinitesimal. The corresponding changes of the generalized vielbein and the dilaton are given by the generalized Lie derivatives
\begin{align}\label{eqn:genliederiv}
  &\delta_\xi E\indices{^A_M} = \mathcal{L}_\xi E\indices{^A_M} =
    \xi^P \partial_P E\indices{^A_M} + 
    (\partial_M \xi^P - \partial^P \xi_M) E\indices{^A_P} 
    \quad \text{and}\\
  &\delta_\xi \phi' = \mathcal{L}_\xi \phi' = \xi^M \partial_M \phi' -
      \frac{1}{2} \partial_M \xi^M \,.
\end{align}
These infinitesimal transformations form the algebra
\begin{equation}
  [\delta_{\xi_1}, \delta_{\xi_2}] = \delta_{\xi_1} \delta_{\xi_2} -
  \delta_{\xi_2} \delta_{\xi_1} = - \mathcal{L}_{[\xi_1, \xi_2]_\mathrm{C}}
\end{equation}
which is governed by the C-bracket
\begin{equation}\label{eqn:Cbracket}
  \left[ \xi_1, \xi_2 \right]_\mathrm{C}^M = \xi_1^N \partial_N \xi_2^M -
    \frac{1}{2} \xi_{1 N} \partial^M \xi_2^N - \left( \xi_1 
    \leftrightarrow \xi_2 \right)\,,
\end{equation}
provided we impose the strong constraint
\begin{equation}\label{eqn:strongconstraint}
  \partial_N \partial^N \cdot = 0
\end{equation}
where $\cdot$ is a place holder for fields, gauge parameters and arbitrary products of them. This is a stronger form of the level-matching constraint $L_0 - \bar L_0 = 0$ of closed string theory. In general this algebra does not satisfy the Jacobi identity and so the generalized diffeomorphisms do not form a Lie group. However, its failure to satisfy the Jacobi identity is of a trivial form that does not generate a gauge transformation on fields satisfying the strong constraint. Thus, it is consistent with the Jacobi identity for symmetry variations on physical fields, which always holds.
A trivial way to solve \eqref{eqn:strongconstraint} is to set $\tilde \partial^i = 0$. In this case, the DFT action \eqref{eqn:dftaction} leads to the NS/NS action \eqref{eqn:nsnsaction} discussed in the introduction. 

\subsection{Equations of motion for the generalized metric}\label{sec:dfteom}
Consistent background solutions of the DFT are obtained by the variation of the DFT action. The variation w.r.t. the generalized metric yields
\begin{equation}  
  \frac{\delta S_\mathrm{DFT}}{\delta \mathcal{H}^{MN}} = \mathcal{K}_{MN}\,.
\end{equation}
This does not lead to the equations of motion for the generalized metric directly, because $\mathcal{H}^{MN}$ is a constrained field. To determine the proper projection that encoded the equations of motion we have to use that the generalized metric is O$(D,D)$ valued and must fulfill
\begin{equation}
  \mathcal{H}^{LM} \eta_{MN} \mathcal{H}^{KN} = \eta^{KL}\,.
\end{equation}
The variation of this constraint leads to
\begin{equation}\label{eqn:constrvariation}
  \delta \mathcal{H}^{LM} \mathcal{H}\indices{^K_M} +
  \mathcal{H}\indices{^L_M} \delta \mathcal{H}^{KM} = 0
\end{equation}
and after some relabeling of indices and using $\mathcal{H}^{ML} \mathcal{H}_{LN} = \delta^M_N$ one obtains
\begin{equation}
  \delta\mathcal{H}^{MN} = -\mathcal{H}^{MK}  
    \delta\mathcal{H}_{KL} \mathcal{H}^{LN}\,.
\end{equation}
As described in \cite{Hohm:2010pp,Hohm:2011si}, the most general variation $\delta\mathcal{H}^{MN}$ satisfying \eqref{eqn:constrvariation} can be written as
\begin{gather}
  \delta\mathcal{H}^{MN} =  \bar{P}^{MK} 
    \delta\mathcal{M}_{KL} P^{LN} +
    P^{MK} \delta\mathcal{M}_{KL} \bar{P}^{LN} \\
    \quad \text{with} \quad \bar{P}^{MN} = \frac{1}{2}\left(\eta^{MN} +
      \mathcal{H}^{MN}\right) \quad \text{and} \quad
      P^{MN} = \frac{1}{2}\left(\eta^{MN} -
      \mathcal{H}^{MN}\right)\,,
\end{gather}
where $\delta \mathcal{M}_{MN}$ is now an arbitrary, unconstrained symmetric variation. Because this new variation is not subject to any constraints, it leads to 
\begin{equation}
  \delta S_\mathrm{DFT} = \int \mathrm{d}^{2D} X \mathcal{K}^{MN} \delta 
    \mathcal{H}_{MN} = \int \mathrm{d}^{2D} X \mathcal{R}_{MN} \delta \mathcal{M}^{MN}\,,
\end{equation}
where
\begin{equation}\label{eqn:genriccitensor}
  \mathcal{R}_{MN} = P_{MK} \mathcal{K}^{KL}
    \bar{P}_{LN} + \bar{P}_{MK} \mathcal{K}^{KL} P_{LN}
\end{equation}
is called the generalized Ricci tensor. Then the equation
\begin{equation}
  \mathcal{R}_{MN}=0
\end{equation}
is the equation of motion for the generalized metric. Because the generalized metric $\mathcal{H}^{MN}$ is symmetric, $\mathcal{K}_{MN}$ and $\mathcal{R}_{MN}$ are symmetric, too. For completeness we give finally the explicit expression for $\mathcal{K}_{MN}$ which arises from the variation of the DFT action with respect to the generalized vielbein\footnote{Within this paper we use the abbreviations
\begin{equation*}
  T_{[a_1 \dots a_n]} = \frac{1}{n!}  \sum\limits_{\sigma\in P} \sign(\sigma) T_{\sigma_1 \dots \sigma_n}
  \quad \text{and} \quad 
  T_{(a_1 \dots a_n)} = \frac{1}{n!}  \sum\limits_{\sigma\in P} 
    T_{\sigma_1 \dots \sigma_n}\,,
\end{equation*}
where $P$ is the set of all permutations of the indices $a_1,\dots,a_n$,
for the (anti)symmetrization of rank $n$ tensors.}:
\begin{gather}
  \mathcal{K}_{MN} = \frac{1}{8} \partial_M \mathcal{H}^{KL} \partial_N \mathcal{H}_{KL}
  - \frac{1}{4}\left( \partial_L - 2(\partial_L \phi')\right) \left( \mathcal{H}^{KL}\partial_K \mathcal{H}_{MN}\right)
  + 2\partial_M\partial_N \phi' \nonumber \\
  - \frac{1}{2} \partial_{(M} \mathcal{H}^{KL} \partial_L \mathcal{H}_{N)K}
  + \frac{1}{2} \left( \partial_L - 2( \partial_L \phi') \right) \left( \mathcal{H}^{KL}\partial_{(M} \mathcal{H}_{N)K} + \mathcal{H}\indices{^K_{(M}} \partial_K \mathcal{H}\indices{^L_{N)}} \right)
  \,.
\end{gather}

\subsection{Covariant formulation of fluxes}\label{sec:covariantfluxes}
Before we discuss how to obtain solutions of the DFT equations of motion, let us connect the DFT background fields to geometric as well as non-geometric fluxes. It will be useful to have an $O(D,D)$ covariant characterization of the fluxes, which combines the geometric and non-geometric fluxes into a single $O(D,D)$ tensor. Without doubling of coordinates, such a description has already been given a few years ago by Ellwood in \cite{Ellwood:2006ya}. There is a straightforward extension of this prescription to DFT, most conveniently in the language of a frame formalism \cite{Siegel:1993th,Hohm:2010xe}. This has been worked out in the recent papers \cite{Aldazabal:2011nj,Aldazabal:2013sca}, giving a slight reformulation of the frame formulation of \cite{Siegel:1993th,Hohm:2010xe} that is somewhat better adapted to the usual description of fluxes. In this formulation the covariant fluxes can be defined covariantly by means of the C-bracket and the $O(D,D)$ inner product as
\begin{equation}\label{eqn:fluxesandparameters}
  \mathcal{F}_{ABC} = \left[ E\indices{_A} , E\indices{_B} 
    \right]_\mathrm{C}^L E_{CL}\,.
\end{equation}
Using the definition of the C-bracket \eqref{eqn:Cbracket}, \eqref{eqn:fluxesandparameters} expands to
\begin{align}\label{eqn:omega_ABCexpanded}
  \mathcal{F}_{ABC} &=  E\indices{_A^N} \partial_N
    E\indices{_B^L} E_{CL} - \frac{1}{2} E_{AN} \partial^L 
    E\indices{_B^N} E_{CL} - ( A \leftrightarrow B ) \nonumber \\
    &= \Omega_{ABC} + \frac{1}{2} \Omega_{CAB} - \Omega_{BAC} -
      \frac{1}{2} \Omega_{CBA} = \Omega_{ABC} + \Omega_{CAB} +
      \Omega_{BCA}\,, 
\end{align}
when introducing the coefficients of anholonomy
\begin{equation}\label{eqn:coeffanholo}
  \Omega_{ABC}=E\indices{_A^N}\partial_N E\indices{_B^M} E_{CM}\,. 
\end{equation} 
They are antisymmetric with respect to its last two indices $B$ and $C$, as a consequence of 
\begin{equation}
  E\indices{_A^N} \partial_N \left( E\indices{_B^M} \eta_{ML} 
    E\indices{_C^L} \right) = E\indices{_A^N} \partial_N \eta_{BC} = 0\,.
\end{equation}
We thus obtain 
\begin{equation}\label{eqn:fluxescoeffanholo}
  \mathcal{F}_{ABC} = \Omega_{ABC} + \Omega_{CAB} + \Omega_{BCA}\,.
\end{equation}
Using the antisymmetric property once more, it is evident that the covariant fluxes are totally asymmetric, 
\begin{equation}
  \mathcal{F}_{ABC} = 3 \Omega_{[ABC]}\,.
\end{equation}
They have three flat indices and thus are subject to double Lorentz transformations. 

For completeness, in the following we explicitly calculate the various components of $\mathcal{F}_{ABC}$ by starting with a generalized vielbein that is `over-parametrized' in the sense that it encodes a two-form $B_{ij}$ and a bi-vector $\beta^{ij}$, as opposed to the physical fields only (i.e., either the two-form or the bivector). Put differently, we have not yet gauge fixed to the physical diagonal subgroup of the double Lorentz group O$({D-1},1)_\mathrm{R}\times$O$(1,{D-1})_\mathrm{L}$ so that there are pure gauge modes left. In a given physical situation one may then gauge fix further to a frame containing only a 2-form, only a bivector, or some intermediate frame. For a gauge without independent $B$-field the covariant fluxes reduce to those identified in \cite{Andriot:2012wx,Andriot:2012an}. Here we give the vielbein with the flat index lowered and the curved one raised:
\begin{equation}
\label{eqn:paramUAM}
  E\indices{_A^M} = \eta_{AB} E\indices{^B_N} \eta^{NM} =
  \begin{pmatrix}
    e\indices{^a_i} + e\indices{^a_j} \beta^{jk} B_{ki} &
    e\indices{^a_j} \beta^{ji}
    \\
    e\indices{_a^j} B_{ji} &
    e\indices{_a^i}
  \end{pmatrix}\,.
\end{equation}
Due to the fact that the covariant fluxes are described by a totally antisymmetric tensor, only 4 of the 8 $D\times D\times D$ blocks ${\mathcal F}_{ABC}$ consists of are independent from each other. Each of these independent blocks, namely ${\mathcal F}_{abc}$, ${\mathcal F}\indices{^a_b_c}$, ${\mathcal F}\indices{^a^b_c}$ and ${\mathcal F}^{abc}$, will now be evaluated. By this calculation, we are able to connect the covariant fluxes with the fluxes $H^{abc}$, $f\indices{^a_b_c}$ (geometric flux), $Q\indices{^a^b_c}$ ($Q$-flux) and $R_{abc}$ ($R$-flux) in flat indices. The three additional fluxes, which were not discussed so far, are common in the description of non-geometric backgrounds. A good overview over their structure and properties is given for example by \cite{Shelton:2005cf, Andriot:2012an}.

We start with ${\mathcal F}_{abc}$ which is given in terms of
\begin{equation}\label{eqn:Fabcfromcoeffanholo}
  {\mathcal F}_{abc} = {\Omega}_{abc} + {\Omega}_{cab} +
    {\Omega}_{bca} = 3{\Omega}_{[abc]}\,.
\end{equation}
Putting \eqref{eqn:paramUAM} into \eqref{eqn:coeffanholo}, the relevant coefficients of anholonomy evaluate to
\begin{equation}
  {\Omega}_{abc}= e\indices{_a^i} e\indices{_b^j} e\indices{_c^k} \left( 
  \partial_i B_{jk} + B_{il} \tilde{\partial}^l B_{jk} \right)\,.
\end{equation}
Combining this result with the antisymmetrization of $\Omega_{ijk}$ in \eqref{eqn:Fabcfromcoeffanholo} gives rise to
\begin{equation}
  {\mathcal F}_{abc} = 3 e\indices{_a^i} e\indices{_b^j} e\indices{_c^k} \left( 
  \partial_{[i} B_{jk]} - B_{l[i} \tilde{\partial}^l B_{jk]} \right) = H_{abc}\,.
\end{equation}
When applying the strong constraint $\tilde \partial_i = 0$, this expression is equivalent to the $H$-flux in flat indices. In the next step, we calculate the three components $\Omega\indices{^a_b_c}$, $\Omega\indices{_a^b_c}$ and $\Omega\indices{_a_b^c}$. These are all combinations with two lowered and one raised index. They are given by the following expressions
\begin{align}\label{eqn:Omega^a_b_c}
  \Omega\indices{^a_b_c} &= e\indices{^a_i} e\indices{_b^j} e\indices{_c^k} \left( \tilde{\partial}^i 
    B_{jk} + \beta^{il} \Omega_{ljk} \right)\;,  \\
  \Omega\indices{_a^b_c} &= e\indices{_a^i} \partial_i e\indices{^b_j} e\indices{_c^j}
    + e\indices{_a^i} B_{ij} \tilde{\partial}^j e\indices{^b_k} e\indices{_c^k}
    + e\indices{_a^i} e\indices{^b_j} e\indices{_c^k} \beta^{jl} \Omega_{ilk}\;,  \\
  \Omega\indices{_a_b^c} &= -\Omega\indices{_a^c_b} \,.
\end{align}
With these three components, the covariant fluxes  ${\mathcal F}\indices{^a_b_c}$ read
\begin{align}
  {\mathcal F}\indices{^a_b_c} &= \Omega\indices{^a_{[b}_{c]}}
  + \Omega\indices{_{[c}^a_{b]}} 
  + \Omega\indices{_{[b}_{c]}^a} = \Omega\indices{^a_{[b}_{c]}} + 
  2 \Omega\indices{_{[c}^a_{b]}} \nonumber \\
  &= 2 \left( e\indices{_{[b}^i} \partial_i e\indices{^a_j} e\indices{_{c]}^j}
    + e\indices{_a^i} B_{ij} \tilde{\partial}^j e\indices{^b_k} e\indices{_c^k} 
    \right) + e\indices{^a_i} e\indices{_b^j} e\indices{_c^k} \left( \tilde{\partial}^i 
    B_{jk} + \beta^{il} H_{ljk} \right)= f^a_{bc}\,.
\end{align}
They are equivalent to the geometric fluxes $f^a{}_{bc}$ in flat indices. This equivalence gets manifest, if a frame is chosen where $\tilde \partial^i = 0$ and $\beta^{ij}=0$ holds. Then $\mathcal{F}\indices{^a_b_c}$ becomes \begin{equation}
  {\mathcal F}\indices{^a_b_c} = 2 e\indices{_{[b}^i} \partial_i e\indices{^a_j} e\indices{_{c]}^j} = f^a_{bc}\,,
\end{equation}
which is exactly the form given by e.g. \cite{Blumenhagen:2013hva}. In order to calculate ${\mathcal F}\indices{^a^b_c}$ one needs the anholonomy coefficient's components
\begin{align}
  \Omega\indices{^a^b_c} &= e\indices{^a_i}\tilde\partial^i e\indices{^b_j} e\indices{_c^j} + e\indices{^a_i} e\indices{^b_j} e\indices{_c^k} \beta^{il} \Omega\indices{_l^j_k} \\
  \Omega\indices{_a^b^c} &= e\indices{_a^i} e\indices{_j^b} e\indices{_k^c} \left( \partial_i \beta^{jk} + B_{il}\tilde\partial^l \beta^{jk} + \beta^{jl} \beta^{km} \Omega_{ilm} \right) \quad \text{and} \\
  \Omega\indices{^a_b^c} &= -\Omega\indices{^a^c_b}\,.
\end{align}
They are combined to
\begin{align}
  \mathcal F\indices{^a^b_c} &=
    \Omega\indices{^{[a}^{b]}_c} +
    \Omega\indices{_c^{[a}^{b]}} + 
    \Omega\indices{^{[b}_c^{a]}} = 2 \Omega\indices{^{[a}^{b]}_c} +
    \Omega\indices{_c^{[a}^{b]}} \nonumber\\
    & = 2 e\indices{^{[a}_i} \tilde\partial^i e\indices{^{b]}_j} e\indices{_c^j} + e\indices{_i^{[a}} e\indices{_j^{b]}} e\indices{_c^k} \left( \partial_k \beta^{ij} + B_{kl}\tilde\partial^l \beta^{ij} -
    \beta^{li} \left[ 2 \Omega\indices{_l^{j}_k}  + \beta^{jn} \Omega_{kln} \right]  \right) = Q_c^{ab}
\end{align}
which is equivalent to the $Q$-flux in flat indices. In the frame $\tilde \partial^i$ and $B_{ij}=0$, this expression transforms into
\begin{equation}
  \mathcal F\indices{^a^b_c} = e\indices{_i^a} e\indices{_j^b} e\indices{_c^k} \left( \partial_k \beta^{ij} - 
    \beta^{l[i} f^{j]}_{kl} \right) =Q^{ab}_c
\end{equation}
and thus is equivalent to the $Q$-flux defined in e.g. \cite{Andriot:2013xca}. Finally, we have
\begin{align}
  \Omega^{abc} &= e\indices{^a_i} e\indices{^b_j} e\indices{^c_k} \left(\tilde\partial^i \beta^{jk} + \beta^{il} \Omega\indices{_l^j^k}\right)\,,
\end{align}
which gives rise to
\begin{equation}\label{eqn:Rflux}
  {\mathcal F}^{abc} = 3 \Omega^{[abc]} =   e\indices{^a_i} e\indices{^b_j} e\indices{^c_k} 3 \left(\tilde\partial^{[i} \beta^{jk]} + \beta^{[il} \partial_l \beta^{jk]} + \beta^{il} B_{ln} \tilde\partial^n \beta^{jk} + \beta^{il} \beta^{jm} \beta^{kn} \mathcal{F}\indices{_l_m_n}\right)
\end{equation}
and is equivalent to the $R$-flux in flat indices. To see this, we use the frame $\tilde \partial^i=0$ and $B_{ij}=0$ in which \eqref{eqn:Rflux} reads
\begin{equation}
  {\mathcal F}^{abc} = e\indices{^a_i} e\indices{^b_j} e\indices{^c_k} 3 \beta^{[il} \partial_l \beta^{jk]} = R^{abc}\,.
\end{equation}
This expression is equivalent to the $R$-flux defined in e.g.\cite{Andriot:2012an}. All these results agree with the ones presented in \cite{Geissbuhler:2013uka,Blumenhagen:2013hva} and show that the covariant fluxes are indeed a generalization of the fluxes known from the SUGRA effective action \eqref{eqn:nsnsaction}.

\section{Twisted backgrounds in DFT}\label{sec:DFTbackgrounds}
When constructing backgrounds for string theory, a major challenge is to find non-trivial solutions for the background field equations. As shown in section~\ref{sec:dfteom}, these equations are derived by varying the DFT action \eqref{eqn:dftaction} with respect to the generalized metric's physical degrees of freedom. As discussed in section~\ref{sec:dfteom}, they are very involved, and in general it is impossible to solve them directly. One way to overcome this problem is to start with known SUGRA solutions, like NS 5-branes or orthogonal intersections of them and apply various T-duality transformations on them \cite{Hassler:2013wsa}. Here we use another technique, namely a \textit{consistent} generalized Scherk-Schwarz compactification. It gives rise to a lower-dimensional effective action which is easier to handle than the full DFT action. This action describes a gauged (super)gravity and is equipped with a scalar potential which considerably restricts the vacua of the effective theory.

Because we use a consistent compactification, the solutions of the effective gauged (super)gravity's field equations can be uplifted to solutions of the DFT background field equations. In fact, the uplift can always be performed in case the background possesses enough isometries. This was discussed e.g. in \cite{Scherk:1978ta,Scherk:1979zr,Pons:2003ka} for standard dimensional reductions of higher dimensional supergravity theories on $(D-d)$-dimensional spaces with $D-d$ isometries.  So in case the generalized Scherk-Schwarz ansatz possesses the doubled number of isometries, i.e. $2(D-d)$ isometries with respect to the coordinates as well as with respect to the dual coordinates, we will argue that the same argument still holds for the consistent uplift of the reduced DFT.

Thus the steps we are performing are summarized by the following diagram:
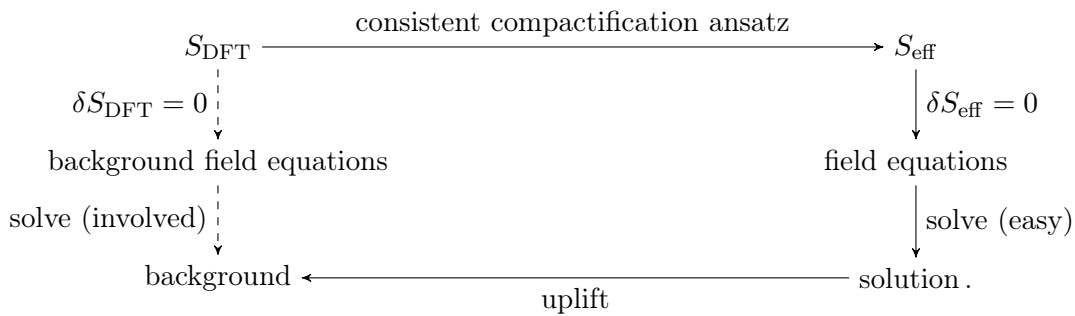
\begin{center}\label{fig:consistentcomp}
  \begin{tikzpicture}[>=stealth',node distance=4em]
    \node (SDFT) {$S_\mathrm{DFT}$};
    \node (Seff) [right of=SDFT, xshift=20em] {$S_\mathrm{eff}$};
    \node (eomeff) [below of=Seff] {field equations};
    \node (soleff) [below of=eomeff] {solution\,.};
    \node (eom) [below of=SDFT] {background field equations};
    \node (sol) [below of=eom] {background};
    \draw[->] (SDFT) -- (Seff) node[midway, above] {consistent compactification ansatz};
    \draw[->] (Seff) -- (eomeff) node[midway, above, anchor=west] {$\delta S_\mathrm{eff} = 0$};
    \draw[->] (eomeff) -- (soleff) node[midway, above, anchor=west] {solve (easy)};
    \draw[->] (soleff) -- (sol) node[midway, below] {uplift} ;
    \draw[->,dashed] (SDFT) -- (eom) node[midway, below, anchor=east] {$\delta S_\mathrm{DFT} = 0$};
    \draw[->,dashed] (eom) -- (sol) node[midway, below, anchor=east] {solve (involved)};
  \end{tikzpicture}
\end{center}
We will now follow the path marked by the solid black lines to find a valid background. The following subsections describe the way from $S_\mathrm{DFT}$ to the solution of the effective field theory's equations of motion. Section~\ref{sec:constrtwists} discussed
 the explicit uplift by considering so called twisted backgrounds, with enough isometries for a consistent uplift.

\subsection{Generalized Kaluza-Klein ansatz}\label{sec:KKansatz}
In every compactification one distinguishes between internal and external, i.e. uncompactified directions. Here we assume that we have $d$ external and $D-d$ internal dimensions. To make this situation manifest, we split the $2D$ components of the vector $X^{\mathcal M}=\begin{pmatrix} \tilde x_i & x^i \end{pmatrix}$ into
\begin{equation}
  X^{\hat M}= \begin{pmatrix} \tilde x_\mu & x^\mu  & Y^M \end{pmatrix} =    
    \begin{pmatrix} \mathds{X} & \mathds{Y} \end{pmatrix}\,, 
    \quad\text{where}\quad \mu=0,\dots, d-1
\end{equation}
counts the external directions and $Y^M$ is an covariant vector in the internal double space. In these conventions the O$(D,D)$ invariant metric \eqref{eqn:etaMN} reads 
\begin{equation}
     \eta_{\hat M\hat N}=\begin{pmatrix}
     0 & \delta^\mu_\nu & 0 \\
      \delta_\mu^\nu & 0 & 0 \\
      0 & 0 & \eta_{MN} 
    \end{pmatrix} \quad \text{and its inverse} \quad
    \eta^{\hat M \hat N}=\begin{pmatrix}
      0 & \delta_\mu^\nu & 0\\
      \delta^\mu_\nu & 0 & 0\\
      0 & 0 & \eta^{MN} 
    \end{pmatrix}\,. 
\end{equation}

In this subsection we will review as warm-up compactifications of DFT, for which the internal $2(D-d)$-dimensional space does not depend on the coordinates in the internal directions. Hence we are basically dealing with compacifications on a doubled torus $T^{2(D-d)}$. Specifically, we demand, that the internal space is invariant under $2(D-d)$ independent isometries. An isometry is a shift of the coordinates $X^{\hat J} \rightarrow X^{\hat J} - K^{\hat J}$ which does not change the generalized metric. Using the generalized Lie derivative, which generates such coordinate shifts, an isometry is defined by
\begin{equation}\label{eqn:killinggenmetric}
  {\mathcal L}_{K^{\hat J}} {\mathcal H}_{\hat M\hat N} = 0\;, 
\end{equation}
where $K^{\hat J}$ is the Killing vector. This is the generalized Killing equations in the generalized geometry of DFT. In total we need $2(D-d)$ independent isometries to construct a consistent compactification ansatz. They are denotes by $K\indices{_I^{\hat J}}$ with $I=1,\dots,2(D-d)$ labeling the different Killing vectors. Condition \eqref{eqn:killinggenmetric} is fulfilled in particular when
\begin{equation}
  {\mathcal L}_{K\indices{_I^{\hat J}}} E\indices{^{\hat A}_{\hat M}} = 0 \quad \rightarrow \quad
  {\mathcal L}_{K\indices{_I^{\hat J}}} {\mathcal H}_{\hat M\hat N} = \left( {\mathcal L}_K E\indices{^{\hat A}_{\hat M}} 
    \right) \delta_{AB} E\indices{^{\hat B}_{\hat N}} + E\indices{^{\hat A}_{\hat M}} 
    \delta_{AB} \left( {\mathcal L}_{K} E\indices{^{\hat B}_{\hat N}} \right) = 0\,, 
\end{equation}
although in general one may impose the weaker condition that the Killing vectors leave the frame field 
invariant only up to a local Lorentz transformation. This equation allows us to use the generalized vielbein $E\indices{^{\hat A}_{\hat M}}$ to look for Killing vectors of the internal space. As a warm up, we begin with the simplest set of Killing vectors namely 
\begin{equation}\label{eqn:KillingvecKK}
  K\indices{_I^{\hat J}} = \begin{pmatrix} 0 & 0 & \delta_I^J \end{pmatrix}\,.
\end{equation}
The corresponding Killing equation then implies that the generalized vielbein $E\indices{^{\hat A}_{\hat M}}$ has to be independent of the internal coordinates $\mathds{Y}$. This condition leads to the constrained vielbein $\widehat{E}\indices{^{\hat A}_{\hat M}}(\mathds{X})$ that depends only on $\mathds{X}$. This implies that the kinetic part of the energy in the $\mathds{Y}$ directions vanishes and the Kaluza-Klein tower of states is consistently truncated to massless states only.

Generalized Lie derivatives on $\widehat{E}\indices{^{\hat A}_{\hat M}}$ should not violate our ansatz by introducing a 
$\mathds{Y}$ dependence. Thus, we restrict the gauge parameters $\xi$ to depend on $\mathds{X}$ only. In the following, $\mathds{Y}$ independent quantities are always marked by a hat. After these restrictions, one is able to decompose the generalized vielbein into several fields which do not mix under generalized diffeomorphisms and the other symmetry transformations in section~\ref{sec:dftandsym}. These fields are
\begin{itemize}
  \item the $d$-dimensional vielbein $e\indices{^\alpha_\mu}$ and
  \item the corresponding $B$-field $B_{\mu\nu}$,
  \item the $\mu=1,\dots,d$ $2(D-d)$-dimensional, covariant vectors $\widehat A_{M\mu}$ and
  \item the O$(D-d,D-d)$ valued vielbein $\widehat E\indices{^A_M}$.
\end{itemize}
They will be considered as the field content of the effective theory which arises after the compactification. Altogether, they completely parameterize the $D^2$ degrees of freedom of the totally gauge fixed generalized vielbein in \eqref{eqn:EAMfixed} and lead to the Kaluza-Klein ansatz
\begin{equation}\label{eqn:vielbeinreparam}
  \widehat E\indices{^{\hat A}_{\hat M}}(\mathds{X}) = \begin{pmatrix}
    e\indices{_\alpha^\mu} & - e\indices{_\alpha^\rho} C_{\mu\rho}
      & - e\indices{_\alpha^\rho} \widehat A_{M\rho} \\
    0 & e\indices{^\alpha_\mu} & 0 \\
    0 & \widehat E\indices{^A_L} \widehat A\indices{^L_\mu} & 
    \widehat E\indices{^A_M}
  \end{pmatrix} \quad \text{with} \quad C_{\mu\nu} = B_{\mu\nu} + 
    \frac{1}{2} \widehat{A}\indices{^L_\mu} \widehat{A}_{L\nu} \,.
\end{equation}
This coincides with the ansatz given in \cite{Hohm:2013nja} once the dependence on internal coordinates is dropped. 
Of course $\widehat E\indices{^{\hat A}_{\hat M}}$ has to be still O$(D,D)$ valued and hence must satisfy \eqref{eqn:vielbeinodd}. This is the case, if and only if
\begin{equation}
  e\indices{_\alpha^\mu} \eta^{\alpha\beta} e\indices{_\beta^\nu} = 
    \eta^{\mu\nu} \quad \text{and} \quad
  \widehat{E}\indices{^A_M} \eta_{AB} \widehat{E}\indices{^B_N} =
    \eta^{MN}\;, 
\end{equation}
i.e., if $\hat{E}$ is  O$(D-d,D-d)$ valued. 

In the $d$ uncompactified space time directions, there are no winding modes. Thus in these directions, the strong constraint \eqref{eqn:strongconstraint} is trivially solved by $\tilde \partial^\mu = 0$ and the partial derivative in doubled coordinates reduces to $\partial^{\hat M} = \begin{pmatrix} \partial_\mu & 0 & \partial^M \end{pmatrix}$. We now compute the action of the generalize diffeomorphisms on the generalized vielbein \eqref{eqn:vielbeinreparam}. They are defined by the generalized Lie derivative \eqref{eqn:genliederiv} with the parameter $\widehat \xi^{\hat M}$. As already mentioned, $\widehat \xi^{\hat M}$ only depend on the coordinates $\mathds{X}$. Its components are 
 \begin{equation}\label{eqn:xiMkaluzaklein}
  \widehat \xi^{\hat M}(\mathds{X}) = \begin{pmatrix} \tilde \xi_\mu & 
    \xi^\mu & \widehat \Lambda^M \end{pmatrix}\,.
\end{equation}
After some algebra, one gets the infinitesimal generalized diffeomorphisms
\begin{align} \label{eqn:genLieKKealphamu}
  {\mathcal L}_{\widehat \xi} e\indices{^\alpha_\mu} &= 
    L_\xi  e\indices{^\alpha_\mu}\,, \\ \label{eqn:genLieKKBmunu}
  {\mathcal L}_{\widehat \xi} B_{\mu\nu} &= L_\xi B_{\mu\nu} + \left(
  \partial_\mu \tilde \xi_\nu - \partial_\nu \tilde \xi_\mu \right) + \partial_{[\mu} \widehat{\Lambda}_M \widehat{A}\indices{^M_{\nu]}}\,, \\
  \label{eqn:genLieKKAMmu}
  {\mathcal L}_{\widehat \xi} \widehat A_{M\mu} &= L_\xi
    \widehat A_{M\mu} - \partial_\mu \widehat \Lambda_M \quad \text{and}\\
  \label{eqn:genLieKKEAM}
  {\mathcal L}_{\widehat \xi} \widehat E\indices{^A_M} &= 
    L_\xi \widehat E\indices{^A_M}
\end{align}
for the various fields of the effective theory, which can also be read off directly from  \cite{Hohm:2013nja}.  
Here, $L_\xi$ is the common Lie derivation in the $d$-dimensional, extended space time. As required, these transformations do not mix different fields. In addition,  they show that the $M=1,\dots,2 (D - d)$ fields $A_{M\mu}$ transform like vectors and the generalized vielbein $\hat E\indices{^A_M}$ transforms like $(D-d)^2$ scalars in the effective theory. Furthermore the vectors posses an abelian U$(1)^{2(D-d)}$ gauge symmetry. This symmetry is generated by the parameters $\widehat{\Lambda}_M$ in \eqref{eqn:genLieKKAMmu}.

With the expressions \eqref{eqn:genLieKKealphamu}--\eqref{eqn:genLieKKEAM} for the generalized Lie derivatives of the various fields, it is immediately clear that the vectors in \eqref{eqn:KillingvecKK} are indeed Killing vectors and thus fulfill
\begin{equation}
  \mathcal{L}_{K\indices{_I^{\hat J}}} e\indices{^\alpha_\mu} =
  \mathcal{L}_{K\indices{_I^{\hat J}}} B_{\mu\nu} =
  \mathcal{L}_{K\indices{_I^{\hat J}}} \widehat{A}_{M\mu} =
  \mathcal{L}_{K\indices{_I^{\hat J}}} \widehat{E}\indices{^A_M} = 0\,. 
\end{equation}

\subsection{Generalized Scherk-Schwarz ansatz}\label{sec:scherkschwarz}
Now we want to deform the Kaluza-Klein ansatz from the previous section. This leads to non-abelian gauge symmetries and massive scalars in the effective theory. Nevertheless, the $2(D-d)$ isometries along the compact internal directions $\mathds{Y}$ shall be kept. In order to achieve this, we replace the $N=1,\dots,2(D-1)$ holonomic basis 1-forms $d Y^N$ of the Kaluza-Klein ansatz with the right-invariant 1-forms \cite{Kaloper:1999yr}
\begin{equation}
  \eta^M = U\indices{^M_N}(\mathds{Y}) d Y^N
\end{equation}
of a Lie group $G$. This is done by the so called twist $U\indices{^N_M}(\mathds{Y})$ and breaks the isometries $G_\mathrm{L} \times G_\mathrm{R}$ of a bi-invariant metric, like the one used in the last section, down to $G_\mathrm{R}$. While $G_\mathrm{R}$ still consists of enough isometries to perform a consistent truncation, $G_\mathrm{L}$ is now used to implement the gauge group of the effective theory. In order to connect this new basis 1-forms with the generalized metric, we have to adapt the scalars $E\indices{^A_M}$ and the vectors $A_{M\mu}$ as
\begin{equation}\label{eqn:twistofgenvielbein}
  E\indices{^A_M}(\mathds{X},\mathds{Y}) = 
    \widehat E\indices{^A_N}(\mathds{X}) U\indices{^N_M}(\mathds{Y})
  \quad \text{and} \quad
  A_{M\mu}(\mathds{X},\mathds{Y}) =
    \widehat A_{N\mu}(\mathds{X}) U\indices{^N_M}(\mathds{Y})\,.
\end{equation}
Of course, one can also write this ansatz in terms of the generalized vielbein
\begin{equation}\label{eqn:twistscherkschw}
  E\indices{^{\hat A}_{\hat M}}(\mathds{X},\mathds{Y}) = 
    \widehat E\indices{^{\hat A}_{\hat N}}(\mathds{X}) 
    U\indices{^{\hat N}_{\hat M}}(\mathds{Y}) 
  \quad \text{with} \quad
  U\indices{^{\hat N}_{\hat M}} = 
  \begin{pmatrix}
    \delta^\mu_\nu & 0 & 0 \\
      0 & \delta_\mu^\nu & 0 \\
      0 & 0 & U\indices{^N_M}
  \end{pmatrix}\,,
\end{equation}
too. As previously emphasised, the generalized vielbein $E\indices{^{\hat A}_{\hat M}}$ has to be O$(D,D)$ valued. The untwisted generalized vielbein $\hat E\indices{^{\hat A}_{\hat M}}$ has this property. Hence the twist $U\indices{^{\hat N}_{\hat M}}$ also has to be O$(D,D)$ valued, which is exactly the case if, and only if, $U\indices{^N_M}$ is O$(D-d,D-d)$ valued.

Dual to the right-invariant 1-forms $\eta^M$ are vectors of the form
\begin{equation}
  \label{eqn:ssansatzxi}
  \xi^{\hat M} = \widehat \xi^{\hat N} U\indices{_{\hat N}^{\hat M}} =
  \begin{pmatrix} \tilde \xi_\mu & \xi^\mu & \Lambda^M \end{pmatrix}\,.
\end{equation}
They generate left-translations acting on $G_\mathrm{L}$. This group, as already explained, was chosen to implement the gauge symmetry of the effective theory. Thus, transformations $\widehat{\xi}^{\hat M}$ with an arbitrary $\mathds{X}$-dependent $\widehat{\xi}^{\hat N}$ represent gauge transformations of the effective theory. To check this, we calculate the generalized Lie derivative of the vector $V_{\hat M} = \widehat{V}_{\hat N} U\indices{^{\hat N}_{\hat M}}$ (which corresponds to a right-invariant 1-form) with the gauge parameter $\xi^{\hat L}$:
\begin{align}
  \mathcal{L}_\xi V_{\hat M} & = \xi^{\hat P} \partial_{\hat P} V_{\hat M} + 
    \left( \partial_{\hat M} \xi^{\hat P} - \partial^{\hat P} \xi_{\hat M} \right)
    V_{\hat P} \nonumber \\
    &= \mathcal{L}_{\widehat \xi} \widehat{V}_{\hat I} U\indices{^{\hat I}_{\hat M}} +
    \widehat{\xi}^{\hat L} \widehat{V}_{\hat N} \left( 
      U\indices{_{\hat L}^{\hat P}} \partial_{\hat P} U\indices{^{\hat N}_{\hat M}} +
      \partial_{\hat M} U\indices{_{\hat L}^{\hat P}} U\indices{^{\hat N}_{\hat P}} - 
      U\indices{^{\hat N}_{\hat P}} \partial^{\hat P} U_{\hat L\hat M} \right) 
    \nonumber \\
    &= \left( \mathcal{L}_{\widehat \xi} \widehat{V}_{\hat I} 
      + \widehat{\xi}^{\hat L} \widehat{V}^{\hat N}
        \left[ \Omega_{\hat L\hat N \hat I} + \Omega_{\hat I\hat L\hat N} -
        \Omega_{\hat N\hat L\hat I} \right] \right) U\indices{^{\hat I}_{\hat M}}
    \nonumber \\ \label{eqn:twistedgenLie}
    &= \left( \mathcal{L}_{\widehat \xi} \widehat V_{\hat I} +
    \mathcal{F}_{\hat I\hat N\hat L} \widehat \xi^{\hat N} \widehat V^{\hat L} \right)
      U\indices{^{\hat I}_{\hat M}}\,.
\end{align}
Here the covariant tensor $\mathcal{F}_{\hat M\hat N\hat L}$ arises through the twist $U\indices{^{\hat M}_{\hat N}}$. A similar deformation of gauge transformations is also part of the DFT formulation of heterotic strings \cite{Hohm:2011ex}. Due to the structure of twist, the covariant tensor vanishes in all external directions $\mathds{X}$. Its non-vanishing components are linked to the covariant fluxes introduced in \eqref{eqn:fluxescoeffanholo} in section~\ref{sec:covariantfluxes} by
\begin{equation}
  \mathcal{F}_{ABC} = \widehat{E}\indices{_A^I}
    \widehat{E}\indices{_B^J} \widehat{E}\indices{_C^K} \mathcal{F}_{IJK}\,.
\end{equation}
Hence in the following we will also call $\mathcal{F}_{IJK}$ covariant fluxes. They are the structure constants of the Lie algebra $\mathfrak{g}_\mathrm{L}$ associated to the Lie group $G_\mathrm{L}$ which we choose as gauge group. Actually,  $G_\mathrm{L}$ is only a group if its associated Lie algebra $\mathfrak{g}_\mathrm{L}$ is consistent, i.e., satisfies the Jacobi identity. Explicit calculations using \eqref{eqn:twistedgenLie} and $\xi_\mu = \tilde \xi^\mu = 0$ show that this condition reads
\begin{equation}\label{eqn:gaugealgclosed}
  \left( \mathcal{F}_{MNL} \mathcal{F}\indices{^L_I_K}
  - \mathcal{F}_{MIL} \mathcal{F}\indices{^L_N_K} \right)
  \widehat{\Lambda}_1^N \widehat{\Lambda}_2^I \widehat V^K = 
  \mathcal{F}_{MNK} \widehat{\Lambda}_{12}^N \widehat V^K\,.
\end{equation}
Thus, covariant fluxes need to fulfill the Jacobi identity
\begin{equation}\label{eqn:quadraticc}
  \mathcal{F}_{LMN}\mathcal{F}\indices{^L_I_K} +
  \mathcal{F}_{LIM}\mathcal{F}\indices{^L_N_K} +
  \mathcal{F}_{LNI}\mathcal{F}\indices{^L_M_K} = 0
    \quad \text{or} \quad 
  \mathcal{F}_{L[MN} \mathcal{F}\indices{^L_{I]}_K} = 0\,, 
\end{equation}
taking the total antisymmetry $\mathcal{F}_{NML} = \mathcal{F}_{[MNL]}$ into account. When \eqref{eqn:quadraticc} holds, we find an effective parameter $\widehat{\Lambda}_{12}^N$ that satisfies \eqref{eqn:gaugealgclosed}, namely
\begin{equation}\label{eqn:structurecoeff}
  \widehat{\Lambda}_{12}^K = \mathcal{F}\indices{^K_I_J} \widehat{\Lambda}_1^I  
    \widehat{\Lambda}_2^J\,. 
\end{equation}
Remembering the fact that the hatted quantities depend only on the extended directions $\mathds{X}$, it becomes clear that the covariant fluxes $\mathcal{F}\indices{^K_I_J}$ may, if at all, also depend only on these directions. Otherwise the gauge algebra would not be closed. But as one sees from \eqref{eqn:coeffanholo},  $\mathcal{F}\indices{^K_I_J}$ depends on the compact directions $\mathds{Y}$ only. So, in order to still close the gauge algebra it have to be
\begin{equation}\label{eqn:constfluxes}
  \mathcal{F}_{NML}\,:\quad  \text{constant}\,.
\end{equation}

The closure condition \eqref{eqn:gaugealgclosed} is known to hold if the strong constraint \eqref{eqn:strongconstraint} is imposed. The strong constraint is satisfied if and only if the twist $U\indices{^M_N}$ also fulfills the strong constraint. But the mapping between covariant fluxes and twists, i.e. the inverse of \eqref{eqn:coeffanholo}, is not trivial. Hence it is not obvious how to impose the strong constraint on the level of the covariant fluxes $\mathcal{F}_{IJK}$ directly. In this context the constraints \eqref{eqn:quadraticc} and \eqref{eqn:constfluxes} are very useful: In case one of them is violated, the strong constraint is violated as well. Another check whether the strong constraint is violated can be performed like this: Provided $\partial_M U\indices{_N^M} = 0$, which we will assume as usual in Scherk-Schwarz compactification, a consequence of the strong constraint is 
\begin{equation}\label{eqn:strongconstfluxes}
  \mathcal{F}_{MNL} \mathcal{F}^{MNL} = 0\,.
\end{equation}
In order to confirm this we compute 
\begin{align}
  \mathcal{F}_{MNL} \mathcal{F}^{MNL} &= 
  3 \Omega_{MNL}\Omega^{MNL} + 6 \Omega_{MNL}\Omega^{LMN} \nonumber \\
  &=  3 \partial_M U\indices{_N^L} \partial^M U\indices{^N_L} - 
  6 \partial_M U\indices{_N^L} \partial_L U^{NM}
  = 3 \partial_M U\indices{_N^L} \partial^M
    U\indices{^N_L} = 0
\end{align}
by using \eqref{eqn:fluxescoeffanholo} and the strong constraint \eqref{eqn:strongconstraint} in the last step. 
To see that the second term in the second line vanishes, we used 
\begin{equation}
  \partial_M \partial_L \left( U\indices{_N^L} U^{NM} \right) = 0 =
    \partial_M U\indices{_N^L} \partial_L U^{NM} \quad \text{with} \quad
    \partial_M U\indices{_N^M} = 0\,.
\end{equation}
The last expression can also be written as
\begin{equation}\label{eqn:FLLN=0}
  U\indices{_L^M} \partial_M U\indices{_N^K} U\indices{^L_K} =  \Omega\indices{_L_N^L} = -\Omega\indices{^L_L_N} = 0 \quad \leftrightarrow \quad \mathcal{F}\indices{^L_L_N} = 0\,.
\end{equation}
A similar condition we will be given below for the Killing vectors. It guarantees that the generalized Lie derivative $\mathcal{L}_{U\indices{_N^M}} \cdot$ leaves densities invariant. Summarizing this discussion, there is the following hierarchy of constraints:
\begin{center}
  \begin{tikzpicture}[>=stealth',node distance=4em]
    \node (sc) [draw, rectangle] {strong constraint $\partial_M\partial^M \cdot = 0$
      and compactification ansatz};
    \node (sum) [below left of=sc, anchor=east, yshift=-2em, draw, rectangle] 
      {$\mathcal{F}_{MNL}\mathcal{F}^{MNL}=0$};
    \node (close1) [below right of=sc, anchor=west, yshift=-1em, draw, rectangle] 
      {$\begin{aligned} \mathcal{F}_{MNL} &=\text{constant} \\
          \mathcal{F}_{L[MN} \mathcal{F}\indices{^L_{I]}_K} &= 0
       \end{aligned}$};
    \node (close2) [below of=close1, anchor=north, yshift=0.5em, draw, rectangle]
      {closure of C-bracket\,.};
      \draw[->] (sc) to node[anchor=east,xshift=-1em] {and $\mathcal{F}\indices{^L_L_N}=0$} (sum);
    \draw[->] (sc) -- (close1);
    \draw[<->] (close1) -- (close2);
  \end{tikzpicture}
\end{center}

Combining \eqref{eqn:twistedgenLie} with \eqref{eqn:genLieKKAMmu} and \eqref{eqn:genLieKKEAM} respectively, one gets the generalized Lie derivatives
\begin{align}
  \label{eqn:gendiffeomorphAMmu}
  {\mathcal L}_\xi A_{M\mu} &= L_\xi A_{M\mu} -\partial_\mu
    \widehat{\Lambda}_M  + \mathcal{F}_{MNL} \widehat{\Lambda}^N 
    A\indices{^L_\mu} \quad \text{and} \\
  \label{eqn:gendiffeomorphEAM}
  {\mathcal L}_\xi E\indices{^A_M} &= L_\xi E\indices{^A_M} +
    \mathcal{F}_{MNL} \widehat{\Lambda}^N E^{AL}\,.
\end{align}
for the twisted fields. It is obvious that both $A_{M\mu}$ and $E\indices{^A_M}$ transform under generalized diffeomorphisms with non-vanishing $\Lambda^M$ as non-abelian vectors fields. With the twist, introduced by the Scherk-Schwarz ansatz, we have transformed the abelian gauge symmetry of the Kaluza-Klein ansatz into a non-abelian one.

The $2(D-d)$ required Killing vectors $K\indices{_I^{\hat J}}$ have to generate right-translations which leave the generalized vielbein $E\indices{^{\hat A}_{\hat M}}$ and also the gauge transformation generated by $\xi^{\hat M}$ invariant. This is the case when
\begin{equation}
  \mathcal{L}_{K\indices{_I^J}} U\indices{_N^M} = 0
\end{equation}
and the $K\indices{_I^{\hat J}}$ in the external directions vanish. 
In this case the generators of  $G_\mathrm{L}$ and $G_\mathrm{R}$ commute
\begin{equation}
  \mathcal{L}_{K\indices{_I^{\hat J}}} \mathcal{L}_{\xi^{\hat M}} \mathcal{H}^{\hat M\hat N} =
    \mathcal{L}_{\xi^{\hat M}} \mathcal{L}_{K\indices{_{\hat I}^{\hat J}}} \mathcal{H}^{\hat M\hat N}\,,
\end{equation}
and one obtains the direct product $G_\mathrm{L}\times G_\mathrm{R}$ from which we started. Of course there are also structure coefficients for the group of isometries associated to the Killing vectors. They are calculated in the same way as the covariant fluxes in \eqref{eqn:twistedgenLie}. This gives rise to
\begin{equation}
  \mathcal{L}_{K\indices{_I^{\hat M}}} K\indices{_J^{\hat N}} = \tilde{\mathcal F}\indices{_I_J^K} K\indices{_K^{\hat N}}\;, 
\end{equation}
with
\begin{equation}\label{eqn:structurekilling}
  \tilde{\mathcal F}\indices{_I_J^K} = 
    K\indices{_I^N} \partial_N K\indices{_J^M} K\indices{^K_M} +
    K^{KN} \partial_N K\indices{_I^M} K_{JM} + 
    K\indices{_J^N} \partial_N K\indices{^K_M} K\indices{_I^M}\,.
\end{equation}
Here $K\indices{^I_J}$ again denotes the inverse transpose of $K\indices{_I^J}$ and $K\indices{_I^L} K\indices{^J_L} = \delta^J_I$. But nevertheless, in general, $K\indices{_I^J}$ is not an O$(D-d,D-d)$ matrix. Hence, its first index cannot be raised or lowered with $\eta_{MN}$ or $\eta^{MN}$, respectively.

Furthermore, the transformations generated by $K\indices{_I^J}$ have to leave densities, like $e^{-2 \phi'}$, invariant. For the Kaluza-Klein ansatz from the last section, this constraint is fulfilled trivially, but here we have to check that
\begin{equation}\label{eqn:invdensity}
  \mathcal{L}_{K\indices{_I^J}} \phi' = K\indices{_I^J} \partial_J \phi' - \frac{1}{2} \partial_J K\indices{_I^J} =
    \frac{1}{2} \partial_J K\indices{_I^J} = 0 \quad \rightarrow \quad \partial_J K\indices{_I^J} = 0\,.
\end{equation}
As for the reset of the paper, we assumed in the first step $\phi'=\,$constant. In analogy with \eqref{eqn:FLLN=0}, this condition can be also expressed in terms of the structure constants $\tilde{\mathcal F}_{IJK}$, namely
\begin{equation}
  \tilde{\mathcal F}\indices{^I_I_J} = 0\,.
\end{equation}

Let us  note that the condition \eqref{eqn:invdensity} can be used to prove that the Lagrangian density does not depend anymore on the internal coordinates. To see this, consider the action of a Killing vector $K_I$ on the Lagrangian defining DFT which, being a scalar density, transforms as 
\begin{equation}\label{eqn:genlieLdft2}
  \delta_{K_I} L_\mathrm{DFT} = \partial_J ( K\indices{_I^J} L_\mathrm{DFT}) = \partial_J K\indices{_I^J} L_\mathrm{DFT} + K\indices{_I^J} \partial_J L_\mathrm{DFT} = K\indices{_I^J} \partial_J L_\mathrm{DFT} = 0\;, 
\end{equation}
where we used \eqref{eqn:invdensity} to drop the term with the partial derivative acting on the Killing vectors $K\indices{_I^J}$. Because $K\indices{_I^J}$ consists of $2(D-d)$ linearly independent vector fields, from this equation we can immediately 
conclude 
\begin{equation}
  \partial_J L_\mathrm{DFT} = 0\,.
\end{equation}
This shows that $L_\mathrm{DFT}$ does not depend on the internal coordinates $\mathds{Y}$ when there are $2(D-d)$ linearly  independent Killing vectors. Hence, according to our notation, the Lagrange density $L_\mathrm{DFT}$ can be written as $\widehat{L}_\mathrm{DFT}$.

\medskip 

In the following we want to argue that the Scherk-Schwarz compactification is consistent in the strong Kaluza-Klein sense that each solution of the lower-dimensional theory can be lifted to a solution of the original, higher-dimensional theory. We first note that, by definition, the Killing vectors leaves the generalized Ricci tensor invariant, 
\begin{align} \label{eqn:genliegenricci}
  \delta_{K_I} \mathcal{R}_{\hat M\hat N} &= 0\,. 
\end{align}
It is now easy to see that this equation is solved by 
\begin{equation}\label{eqn:ansatzgenricci}
  \mathcal{R}_{\hat M\hat N} = U\indices{^{\hat I}_{\hat M}} 
    \widehat{\mathcal R}_{\hat I\hat J} U\indices{^{\hat J}_{\hat N}}\,,
\end{equation}
using 
\begin{equation}
  \mathcal{L}_{K_I} U\indices{^{\hat L}_{\hat M}} = 0 \quad \text{and} \quad
  \mathcal{L}_{K_I} \widehat{\mathcal R}_{\hat L \hat K} = 0\,.
\end{equation}
Now, acting with $U\indices{_{\hat I}^{\hat M}}$, the inverse transpose of $U\indices{^{\hat I}_{\hat M}}$, we can conclude 
\begin{equation}
  \mathcal{R}_{\hat M\hat N} = 0 \quad \leftrightarrow \quad \widehat{\mathcal R}_{\hat M\hat N} = 0\,.
\end{equation}
Hence, once the $\mathds{Y}$-independent part of the equations of motion is solved we can immediately construct the higher-dimensional Ricci tensor (satisfying the original DFT equations) via \eqref{eqn:ansatzgenricci}, thus showing the consistency of the Scherk-Schwarz reduction. Put differently, the dashed and the solid path in the diagram on page \pageref{fig:consistentcomp} commute. For our analysis in subsequent chapters we need the explicit definition of the Ricci tensor in the lower-dimensional theory,which is computed from
\begin{equation}
  \widehat{\mathcal K}_{\hat M\hat N} = \frac{\delta \widehat{S}_\mathrm{eff}}{\delta \widehat{\mathcal H}^{\hat M\hat N}}
    \quad \text{with} \quad S_\mathrm{eff}=\int d^{2D} X \,\widehat{L}_\mathrm{DFT}
\end{equation}
using the projection
\begin{equation}
  \widehat{\mathcal R}_{\hat M\hat N} =  \widehat{P}_{\hat M\hat K} \widehat{\mathcal K}^{\hat K\hat L}
  \bar{\widehat{P}}_{\hat L\hat N} + \bar{\widehat{P}}_{\hat M\hat K} \widehat{\mathcal K}^{\hat K\hat L} \widehat{P}_{\hat L\hat N}\;. 
\end{equation}
(See section~\ref{sec:dfteom} for details on the projection). 

Finally, we want to mention, that the generalized fluxes presented in this section are closely related to the embedding tensor $\Theta\indices{_I^\alpha}$ of gauged supergravities. In this context they describe a subset of the global O$({D-d},{D-d})$ symmetry transformations of the compact directions, which is promoted to a gauge symmetry in the effective theory. Comparing the formalism reviewed in \cite{Samtleben:2008pe} and the one shown here, one finds the connection
\begin{equation}\label{eqn:embeddingtensor}
  \mathcal{F}\indices{_I_J^K} = \Theta\indices{_I^\alpha} 
    (t_\alpha)\indices{_J^K} = \left( X_I \right)\indices{_J^K} \,,
\end{equation}
where $t_\alpha$ are $(D-d)\left[2(D-d) - 1\right]$ different O$({D-d},{D-d})$ generators and $(t_\alpha)\indices{_J^K}$ is the corresponding representation with respect to $2(D-d)$-dimensional vectors. One imposes two consistency constraints on the embedding tensor, namely the linear and the quadratic constraint. An explicit discussion of these constrains for $D-d=2,\,3$ and the connection to DFT is given in \cite{Dibitetto:2012rk}.

\subsection{Gauged (super)gravity and its vacua}\label{sec:gaugedgravity}
In section~\ref{sec:scherkschwarz}, we proved that a consistent Scherk-Schwarz ansatz leads to an $\mathds{Y}$-independent effective action $S_\mathrm{eff}$. The effective action is most conveniently obtained by starting from the formulation in \cite{Hohm:2013nja}, which reduces to the previous results in \cite{Aldazabal:2011nj,Geissbuhler:2011mx} for a Scherk-Schwarz ansatz. Following \cite{Hohm:2013nja}, let us first define a derivate
\begin{equation}\label{eqn:covderiv}
  D_\mu = \partial_\mu - \mathcal{L}_{A\indices{^M_\mu}}
\end{equation}
which transforms covariantly under gauge transformations \eqref{eqn:ssansatzxi}. Applied on the generalized metric $\mathcal{H}_{MN}$, it gives rise to
\begin{align}
  D_\mu \mathcal{H}_{MN} &= U\indices{^I_M} \widehat{D}_\mu \widehat{\mathcal H}_{IJ} U\indices{^J_N} \quad
  \text{with} \nonumber \\
  \widehat{D}_\mu \widehat{\mathcal H}_{MN} &= \partial_\mu 
    \widehat{\mathcal H}_{MN}
  + \mathcal{F}\indices{_M_J^I} \widehat{A}\indices{^J_\mu} 
    \widehat{\mathcal H}_{IN}
  + \mathcal{F}\indices{_N_J^I} \widehat{A}\indices{^J_\mu} 
    \widehat{\mathcal H}_{MI}\,.
\end{align}
The field strength of the gauge field $A\indices{_\mu^M}$ is defined in analogy with Yang-Mills theory by setting
\begin{align}\label{eqn:F_mu_nu^M}
  F\indices{^M_\mu_\nu} &= 2 \partial_{[\mu} A\indices{^M_{\nu]}} - [ A_\mu, A_\nu ]^M_\mathrm{C} = \widehat{F}\indices{^N_\mu_\nu} U\indices{_N^M} \quad \text{with} \nonumber\\
  \widehat{F}\indices{^M_\mu_\nu} &= 2 \partial_{[\mu}
  \widehat{A}\indices{^M_{\nu]}} -
    \mathcal{F}\indices{^M_N_L} \widehat{A}\indices{^N_\mu}
    \widehat{A}\indices{^L_\nu}\,.
\end{align}
It describes how two covariant derivative commute
\begin{equation}
  [D_\mu, D_\nu] = - \mathcal{L}_{F\indices{^M_\mu_\nu}}\,.
\end{equation}
As shown in \cite{Hohm:2013nja}, $F\indices{^M_\mu_\nu}$ in general does not transform covariantly under gauge transformations,
\begin{equation}
  \Delta_\xi F\indices{_\mu_\nu^M}  = \delta_\xi F\indices{_\mu_\nu^M} - \mathcal{L}_\xi F\indices{_\mu_\nu^M} = \partial^M ( \partial_{[\mu} \xi^N A_{\nu]N} )\,.
\end{equation}
This problem is fixed by adding the partial derivative of a 2-form gauge potential to the field strength defined in \eqref{eqn:F_mu_nu^M} which compensates for the wrong transformation behavior. But due to the special properties of the Scherk-Schwarz ansatz for fields \eqref{eqn:twistofgenvielbein} and gauge parameter \eqref{eqn:ssansatzxi}, the failure of covariance vanishes because the expression in the bracket depends on the external directions only. Hence for a Scherk-Schwarz compactification, $F\indices{_\mu_\nu^M}$ is already a covariant field strength. A short calculation, where the result (2.32) from \cite{Hohm:2013nja} is used, shows that also the Bianchi identity
\begin{equation}
  D_{[\mu} F\indices{^M_\nu_{\rho]}} = 0 
\end{equation}
is fulfilled for $F\indices{^M_\mu_\nu}$. Let us next discuss the field strength for the $B$-field, which is extended by a CS terms in order to be invariant under gauge transformations. This gives rise to the field strength
\begin{equation}
  \widehat{G}_{\mu\nu\rho} = 
  3\partial_{[\mu} B_{\nu\rho]} + 3\partial_{[\mu} 
    \widehat{A}\indices{^M_\nu}
    \widehat{A}_{M \rho ]} - \mathcal{F}_{MNL} 
    \widehat{A}\indices{^M_\mu}
    \widehat{A}\indices{^N_\nu}
    \widehat{A}\indices{^L_\rho}\,.
\end{equation}
It transforms covariantly and fulfills the Bianchi identity
\begin{equation}
  \partial_{[\mu} G_{\nu\rho\lambda]} = 0\,.
\end{equation}

With these quantities at hand, the Kaluza-Klein action in \cite{Hohm:2013nja} reads
\begin{gather}
  S_\mathrm{eff} = \int \mathrm{d}x^{(D-d)} \sqrt{-g} e^{-2\phi} \Bigl( \mathcal{R} + 
    4 \partial_\mu\phi \partial^\mu\phi -\frac{1}{12}  \widehat{G}_{\mu\nu\rho} \widehat{G}^{\mu\nu\rho} \nonumber \\
    \label{eqn:ddimeffaction} \qquad
    -\frac{1}{4} \widehat{\mathcal H}_{MN} \widehat{F}^{M\mu\nu} \widehat{F}\indices{^N_\mu_\nu}
    +\frac{1}{8} \widehat{D}_\mu \widehat{\mathcal H}_{MN} \widehat{D}^\mu \widehat{\mathcal H}^{MN}
    -\widehat{V}\Bigr)\,.
\end{gather}
Here $\mathcal{R}$ denotes the scalar curvature in the external directions. In the internal directions, the Lagrange density $L_\mathrm{DFT}$ is constant. Thus the integrals in these direction can be solve and give rise to a global factor, which is neglected in \eqref{eqn:ddimeffaction}. This result is equivalent to the one presented by \cite{Aldazabal:2011nj}. Finally on has to calculate the scalar potential
\begin{equation}
  \widehat{V} = - \widehat{R}(\phi', \widehat{\mathcal H}^{MN})\,.
\end{equation}
Due to the properties of the Scherk-Schwarz ansatz, it is constant with respect to the internal direction $\mathds{Y}$. Hence it is sufficient to calculate it at one special point, lets say $Y^N=0$. Using the definition \eqref{eqn:genricciscalar}, $\phi'=\text{const.}$,
\begin{equation}\label{eqn:diUY=0}
  \left. \partial_I U\indices{^J_K} \right|_{Y^N=0} = \Omega\indices{_I^J_K} \quad \text{and} \quad
  \left. \partial_I \partial_J U\indices{^L_K} \right|_{Y^N=0} = \Omega\indices{_{(I}^L_M}
  \Omega\indices{_{J)}^L_K}\,,
\end{equation}
one obtains after some algebra
\begin{equation}\label{eqn:scalarpotential}
  \widehat V = -\frac{1}{4} \mathcal{F}\indices{_I^K^L}
    \mathcal{F}\indices{_J_K_L} \widehat{\mathcal H}^{IJ} +
    \frac{1}{12} \mathcal{F}_{IKM}
    \mathcal{F}_{JLN} \widehat{\mathcal H}^{IJ} \widehat{\mathcal H}^{KL} 
    \widehat{\mathcal H}^{MN}\,.
\end{equation}
Again, this result is consistent with \cite{Aldazabal:2011nj,Geissbuhler:2013uka}. In the remaining part of this section and in section~\ref{sec:minkowski} all quantities belong to the effective theory and thus only depend on the $d$ external coordinates $\mathds{X}$. To avoid overloading the notation there, we drop the hat we introduced to emphasis that quantities depend on $\mathds{X}$ only. In section~\ref{sec:constrtwists}, we start to use the hat to distinguish between $\mathds{X}$ and $\mathds{Y}$ dependent quantities again.

Since we have performed a consistent compactification, each solution of the effective action is also a solution of the DFT we started with. So in order to find consistent backgrounds we have to solve the field equations of the effective action. These equation are obtained by the variation of the effective action $S_\mathrm{eff}$ which gives rise to
\begin{align}
  \label{eqn:eomeffmetric}
  0 &= \frac{\delta S_\mathrm{NS}}{\delta g^{ij}} - 
    \frac{1}{2} \mathcal{H}_{MN} F\indices{^M_\mu^\rho} F\indices{^N_\nu_\rho}
    + \frac{1}{8} D_\mu \mathcal{H}_{MN} D_\nu \mathcal{H}^{MN} \\
  0 &= \frac{\delta S_\mathrm{NS}}{\delta \phi} - \frac{1}{4} 
    \mathcal{H}_{MN} F^{M\mu\nu} F\indices{^N_\mu_\nu} + \frac{1}{8} 
      D_\mu \mathcal{H}_{MN} D^\mu \mathcal{H}^{MN} - V \\
  0 &= 2 D_\nu \left( \mathcal{H}_{MN} F^{N\mu\nu} \right) - 4\partial_\nu \phi \mathcal{H}_{MN} F^{N\mu\nu} + F\indices{^M_\nu_\rho} G^{\mu\nu\rho} + \mathcal{F}\indices{_M_N^L} \mathcal{H}_{LK} D^\mu \mathcal{H}^{NK}
  \quad \text{and} \\ \label{eqn:eomgenRicci}
  0 &= P_{MK} \mathcal{K}^{KL}
    \bar{P}_{LN} + \bar{P}_{MK} \mathcal{K}^{KL} P_{LN}
\end{align}
with
\begin{equation} \label{eqn:eomkappaMN}
  \mathcal{K}^{MN} = F^{M\mu\nu} F\indices{^N_\mu_\nu} +
    D_\mu D^\mu \mathcal{H}^{MN} - 2\partial_\mu \phi D^\mu \mathcal{H}^{MN}
+ 4 \frac{\delta V}{\delta \mathcal{H}_{MN}}
\end{equation}
and additionally, the well know equations of motion for the string's $NS/NS$ sector
\begin{align}
  \frac{\delta S_\mathrm{NS}}{\delta g^{ij}} &=
    \mathcal{R}_{\mu\nu} + 2\nabla_\mu\partial_\nu \phi 
    - \frac{1}{4} G_{\mu\rho\lambda} G\indices{_\nu^\rho^\lambda}
    \\
  \frac{\delta S_\mathrm{NS}}{\delta \phi} &=
    \mathcal{R} + 4 \left( \nabla_\mu \nabla^\mu \phi - 
    \partial_\mu \phi \partial^\mu \phi \right) 
    - \frac{1}{12} G_{\mu\nu\rho} G^{\mu\nu\rho} \\
  0 &= \nabla^\mu G_{\mu\nu\rho} - 2 \partial^\mu \phi  G_{\mu\nu\rho}
\end{align}
in the low energy approximation. In \eqref{eqn:eomgenRicci} and \eqref{eqn:eomkappaMN}, we have applied the projectors discussed in section~\ref{sec:dfteom}. They respect that not all components of ${\mathcal H}^{MN}$ are physical degrees of freedom.

\section{Minkowski vacua}\label{sec:minkowski}
There are various possibilities how the solve the equations of motion \eqref{eqn:eomeffmetric}-\eqref{eqn:eomkappaMN} of the effective theory. The most straightforward one is to assume that we have a $d$-dimensional Minkowski space. In this case the metric is $g_{\mu\nu}=\eta_{\mu\nu}$ while the dilaton $\phi$ and the generalized metric $\mathcal{H}^{MN}$ of the internal space are constant. Furthermore the $B$-field $B_{\mu\nu}$ and the vectors $A_{M\mu}$ vanish. Now the field equations, discussed in the last section, simplify dramatically into
\begin{equation}\label{eqn:minkowskivacuum}
  \mathcal{R}_{\mu\nu} = 0\,, \quad V = 0 \quad \text{and} \quad 
  \mathcal{K}^{MN} = \frac{\delta V}{\delta \mathcal{H}_{MN}}\,.
\end{equation}
The vacua obtained by these equations fulfill the following requirements:
\begin{itemize}
  \item They correspond to minima of the effective gauged supergravity potential that must have vanishing cosmological constant. Hence the uncompatified dimensions are described by flat Minkowski space time. At this point it is worth noting that the generalized curvature $\cal R$ of DFT in the internal directions $\mathds{Y}$ precisely corresponds to the vacuum energy in the effective theory. Hence the vanishing of the generalized Ricci tensor ${\cal R}_{MN}$ ensures that we are dealing with vacua with vanishing cosmological constant.
 \item The fluctuations around the Minkowski vacua are stable, i.e. the scalar mass matrix is at lest positive semi-definite, as we show in section~\ref{sec:spectrum}. Hence, the scalar potential in general leads to the stabilization of some moduli.
\end{itemize}

In order to solve the equations \eqref{eqn:minkowskivacuum}, let us fist have a closer look at the variation of the scalar potential \eqref{eqn:scalarpotential} with respect to the generalized metric, 
\begin{equation}\label{eqn:kappamnfull}
  \mathcal{K}^{MN} = \frac{\delta V}{\delta \mathcal{H}_{MN}} =
    \frac{1}{4}\left( -\mathcal{F}^{MKL} 
    \mathcal{F}\indices{^N_K_L} + \mathcal{F}\indices{^M_I_K} \mathcal{F}\indices{^N_J_L} 
    \mathcal{H}^{IJ} \mathcal{H}^{KL} \right)\,.
\end{equation}
It has to be evaluated for the value $\bar{\mathcal H}^{MN}$, which $\mathcal{H}^{MN}$ acquires for the vacuum. We express this value in terms of the vacuum's generalized vielbein
\begin{equation}\label{eqn:vacuumgenvielbein}
  \bar{\mathcal H}^{MN} = \bar{E}\indices{_A^M} \delta^{AB} \bar{E}\indices{_B^N}\,.
\end{equation}
In the following, flat and curved indices will be related by means of this \textit{background} frame field, which in particular has the consequence that objects with flat indices are $\mathds{X}$-dependent that usually are constant. By applying this prescription to the indices of \eqref{eqn:kappamnfull}, one obtains
\begin{equation}\label{eqn:confluxesintern} 
  \mathcal{K}^{MN} = \frac{1}{4}\left( \mathcal{F}\indices{^M_A_B} \eta^{BC}
    \mathcal{F}\indices{^N_C_D} \eta^{DA} - \mathcal{F}\indices{^M_A_B} \delta^{BC}
      \mathcal{F}\indices{^N_C_D} \delta^{DA} \right)\,.
\end{equation}
A further simplification is achieved when barred indices are used (see \eqref{eqn:trafobaredind} in section~\ref{sec:dftandsym}). In this case the invariant metric $\eta_{\bar A\bar B}$ and the flat generalized metric $\delta_{\bar A\bar B}$ have non-vanishing entries for $\bar A=\bar B$ only. Using this simplification one is able to explicitly evaluate the two terms in \eqref{eqn:confluxesintern} ($\sigma=-1$ gives rise to the first term, while $\sigma=+1$ reproduces to the second one) as
\begin{equation}
\begin{split}
  \mathcal{F} \indices{^{\bar M}^{\bar a}^{\bar b}}
  \mathcal{F} \indices{^{\bar N}^{\bar c}^{\bar d}}
  \eta_{\bar b\bar c} \eta_{\bar a\bar d}
  + 2\sigma 
  \mathcal{F}\indices{^{\bar M}^{\bar a}_{\bar b}}
  \mathcal{F}\indices{^{\bar N}_{\bar c}^{\bar d}}
  \eta^{\bar b\bar c} \eta_{\bar a\bar d}
  \,+\,& \mathcal{F}\indices{^{\bar M}_{\bar a}_{\bar b}}
  \mathcal{F}\indices{^{\bar N}_{\bar c}_{\bar d}}
  \eta^{\bar b\bar c} \eta^{\bar a\bar d} \\
  &=
  \begin{cases}
    \mathcal{F}\indices{^{\bar M}_{\bar A}_{\bar B}} \eta^{\bar B\bar C} 
      \mathcal{F}\indices{^{\bar N}_{\bar C}_{\bar D}} \eta^{\bar D\bar A} &
      \text{for } \sigma=-1 \\ 
    \mathcal{F}\indices{^{\bar M}_{\bar A}_{\bar B}} \delta^{\bar B\bar C}
      \mathcal{F}\indices{^{\bar N}_{\bar C}_{\bar D}} \delta^{\bar D\bar A} &
      \text{for } \sigma=+1
  \end{cases}\,,
 \end{split} 
\end{equation}
where we have used the parameterization
\begin{equation}
  \mathcal{F}_{\bar M\bar A\bar B} =
    \begin{pmatrix}
      \mathcal{F}\indices{_{\bar M}^{\bar a}^{\bar b}} &
      \mathcal{F}\indices{_{\bar M}^{\bar a}_{\bar b}} \\
      \mathcal{F}\indices{_{\bar M}_{\bar a}^{\bar b}} &
      \mathcal{F}\indices{_{\bar M}_{\bar a}_{\bar b}}
    \end{pmatrix}
\end{equation}
for the covariant fluxes. With this result it is straightforward to compute
\begin{equation}
  \mathcal{K}_{\bar M\bar N} =   
    \mathcal{F}\indices{^{\bar M}_{\bar b}^{\bar a}}
    \mathcal{F}\indices{^{\bar N}_{\bar c}^{\bar d}}
    \eta^{\bar b\bar c} \eta_{\bar a\bar d}\,.
\end{equation}
Furthermore the projectors $P_{MK}$ and $\bar{P}_{LN}$, needed to calculate the generalized Ricci tensor \eqref{eqn:genriccitensor}, take the simple form
\begin{equation}
  \bar{P}_{\bar A\bar B} = \frac{1}{2}\left( \eta_{\bar A\bar B} + 
    \delta_{\bar A\bar B} \right) = \begin{pmatrix}
      0 & 0 \\ 0 & \eta_{\bar a\bar b} \end{pmatrix} \quad \text{and} 
      \quad
  P_{\bar A\bar B} = \frac{1}{2}\left( \eta_{\bar A\bar B} -
    \delta_{\bar A\bar B} \right) = \begin{pmatrix}
      \eta_{\bar a\bar b} & 0 \\ 0 & 0 \end{pmatrix} \;, 
\end{equation}
in barred, flat indices. Hence the generalized Ricci scalar reads
\begin{equation}
  \mathcal{R}_{\bar A\bar B} = - \begin{pmatrix}
    0 & \mathcal{K}\indices{^{\bar a}_{\bar b}} \\
    \mathcal{K}\indices{_{\bar a}^{\bar b}} & 0
  \end{pmatrix}\,.
\end{equation}
This tensor is symmetric and thus the equation of motion $\mathcal{R}_{MN}=0$ reduces to
\begin{equation}\label{eqn:fluxeseom}
  \mathcal{K}\indices{^{\bar a}_{\bar b}} =
    \mathcal{F}\indices{^{\bar a}_{\bar d}^{\bar c}}
    \mathcal{F}\indices{_{\bar b}_{\bar e}^{\bar f}}
    \eta^{\bar d\bar e} \eta_{\bar c\bar f} 
  = 0\,.
\end{equation}
Only backgrounds that  satisfy this equation are consistent. Thus in addition to \eqref{eqn:quadraticc} and \eqref{eqn:constfluxes}, we have to impose the further constraint \eqref{eqn:fluxeseom} on the generalized fluxes. Like the Jacobi identity \eqref{eqn:quadraticc}, it is quadratic in the fluxes. 

In summary, a valid background (without warp factor) is the direct product of a $d$-dimen\-sional Minkowski space and a twisted torus in the compact $(D-d)$-dimensional space. The twist of the torus is described in terms of the covariant fluxes $\mathcal{F}_{ABC}$. They are not arbitrary, but severely constrained.

\subsection{Spectrum of the effective theory}\label{sec:spectrum}
In the last section we discussed vacua for the effective field theory in $d$ dimensions. Now the focus is on small perturbations around these vacua. They play an important r\^ole in the process of moduli stabilization, which fixes some or even all of the scalar fields $\mathcal{H}_{MN}$. This process is governed by mass terms in the effective field theory's Lagrangian. Due to these terms some scalars obtain masses and are not excited in the ground state.

The mass term arises from the second order variation of the scalar potential, 
\begin{equation}\label{eqn:secondvarV}
  \delta^2 V = \sum\limits_{\alpha\,,\,\beta}\left(
    \frac{\delta^2 V}{\delta \mathcal{H}_{IJ} \delta \mathcal{H}_{KL}}
    \frac{\delta \mathcal{H}^{IJ}}{\delta \phi_\alpha}
    \frac{\delta \mathcal{H}^{KL}}{\delta \phi_\beta}
      +
  \frac{\delta V}{\delta \mathcal{H}_{KL}}
    \frac{\delta^2 \mathcal{H}^{KL}}{ \delta \phi_\alpha \, 
    \delta \phi_\beta} \right) \delta \phi_\alpha \delta
    \phi_\beta \,.
\end{equation}
Here we have taken into account that $\mathcal{H}^{MN}$ has to be O$({D-d},{D-d})$ valued and thus not all of its $2(D-d)(D-d-1)$ entries correspond to physical degrees of freedom. So we express the generalized metric $\mathcal{H}^{MN}$ in terms of scalar fields $\phi_\alpha$, $\alpha = 1,\dots,(D-d)^2$, which correspond to unconstrained, physical degrees of freedom. Furthermore, we define
\begin{equation}
  \mathcal{M}_{IJKL} = \frac{\delta^2 V}{\delta \mathcal{H}_{IJ} \delta \mathcal{H}_{KL}} =
\frac{1}{2} {\mathcal F}_{IKM}
  {\mathcal F}_{JLN} {\mathcal H}^{MN}
\end{equation}
in analogy with \eqref{eqn:kappamnfull} and use the abbreviation
\begin{equation}
  \left( h_\alpha \right){}^{IJ} = 
    \frac{\delta {\mathcal H}^{IJ}}{\delta \phi_\alpha} \,.
\end{equation}
Now, \eqref{eqn:secondvarV} takes the form
\begin{equation}\label{eqn:secondvarV2}
  \delta^2 V = \sum\limits_{\alpha\,,\,\beta} \left[
    \mathcal{M}_{IJKL} \left( h_\alpha 
    \right){}^{IJ} \left( h_\beta
    \right){}^{KL} + \mathcal{K}_{KL} \frac{\delta }{\delta \phi_\alpha} 
    \left( h_\beta \right){}^{KL} \right]
    \delta \phi_\alpha \delta \phi_\beta \,.
\end{equation}
One can regard $\left( h_\alpha \right){}^{IJ}$ as an infinitesimal generator of a field variation of ${\cal H}^{IJ}$. 
Thus it has to be compatible with the constraint \eqref{eqn:constrvariation}. It is convenient to work in flat indices like in \eqref{eqn:confluxesintern}. We again use the generalized vielbein $\bar{E}\indices{^A_M}$ of the vacuum to transform curved indices into flat ones. Then the constraint \eqref{eqn:constrvariation} on the variation reads 
\begin{equation}\label{eqn:constvariation2}
  \left( h_\alpha \right){}^{AC} \eta_{CD} \delta^{DB} + 
    \delta^{AC} \eta_{CD} \left( h_\alpha \right){}^{DB} = 0\,.
\end{equation}
In order to construct all generators which fulfill this equation, we switch to barred indices and define
\begin{equation}
  \left( h_{\bar A\bar B} \right)^{\bar C\bar D} =
    \sqrt{2} \delta^{\bar C}_{[\bar A} \delta_{\bar B]\bar E}
    \eta^{\bar E\bar D} \quad \text{with} \quad \alpha = \begin{pmatrix} \bar A & \bar B \end{pmatrix}\,.
\end{equation}
For $\bar A < \bar B$ this leads to $2(D-d)(D-d-1)$ independent generators. Only $(D-d)^2$ are symmetric, the others are antisymmetric. We drop the antisymmetric ones, because the generalized metric is symmetric and so are its variations. Finally we switch back to unbarred indices. With these generators at hand, the generalized metric can be expressed by the exponential map
\begin{equation}\label{eqn:genmetricfluctuation}
  \mathcal{H}^{AB} = \prod\limits_\alpha \exp\left[ \sum\limits_\alpha \left(
    h_\alpha \right)^{AB} \phi_\alpha \right]  = 
  \delta^{AB} + \sum\limits_\alpha (h_\alpha)^{AB} \phi_\alpha +
  \frac{1}{2} \sum\limits_{\alpha,\,\beta} (h_\alpha)^{AC} \delta_{CD} 
    (h_\beta)^{DB} \phi_\alpha \phi_\beta + \dots\;.
\end{equation}
We recall that we have used the vacuum vielbein to flatten curved indices. In the vacuum, all $\phi_\alpha$ vanish and according to \eqref{eqn:genmetricfluctuation}, the generalized metric equals $\mathcal{H}^{AB} = \delta^{AB}$. Back in curved indices this gives rise to the vacuum generalized metric $\bar{\mathcal H}^{MN} = \mathcal{H}^{MN}(\phi_\alpha=0)$. With the parameterization of the generalized metric in \eqref{eqn:genmetricfluctuation}, one obtains
\begin{equation}
  \left. \frac{\delta^2}{\delta \phi_\alpha \, \delta \phi_\beta}
    \mathcal{H}^{AB} \right|_{\phi_\gamma = 0}
    = \begin{cases}
        \left( h_\alpha \right)^{AC} \delta_{CD} \left( h_\beta
        \right)^{DB} & \text{for } \alpha \le \beta \\
        \left( h_\beta \right)^{AC} \delta_{CD} \left( h_\alpha
        \right)^{DB} & \text{otherwise}
    \end{cases}\,.
\end{equation}
Using this result and
\begin{equation}  
  \mathcal{H}^{MN} = \bar{E}\indices{_A^M} \mathcal{H}^{AB} \bar{E}\indices{_B^N}\,,
\end{equation}
one is able to evaluate the variation \eqref{eqn:secondvarV} explicitly. Finally, \eqref{eqn:secondvarV2} gives rise to 
\begin{equation}
  \delta^2 V = \sum\limits_{\alpha\,,\,\beta} M_{\alpha\beta}
    \delta\phi_\alpha \delta\phi_\beta
\end{equation}
with the symmetric mass matrix
\begin{equation}\label{eqn:massmatrix}
  M_{\alpha\beta} = \left( {\mathcal M}_{ABCD} + 
    {\mathcal K}_{AD} \delta_{BC} \right) 
    \left( h_\alpha \right){}^{AB} \left( h_\beta \right){}^{CD}\,.
\end{equation}
In order to identify massive scalars excitations, this matrix has to be diagonalized. Because $M_{\alpha\beta}$ is symmetric, this is always possible and leads to $(D-d)^2$ eigenvalues $\lambda_\alpha$ and the corresponding, orthonormal eigenvectors $v_\alpha$ with the components $\left(v_\alpha\right)_\beta$. In order to diagonalize 
we rotate the generators $\left( h_\alpha \right){}^{AB}$ by defining
\begin{equation}\label{eqn:rotatedgenh}
  ({\bar h}_\alpha ){}^{AB} := \sum\limits_{\beta}
    \left(v_\alpha\right)_\beta \left( h_\beta \right){}^{AB}\,.
\end{equation}
The generalized metric $\mathcal{H}^{AB}$ in \eqref{eqn:genmetricfluctuation} has to invariant under this rotation. Thus one also has to rotate the scalar fields
\begin{equation}
  {\bar\phi}_\alpha := \sum\limits_{\beta} \left(v_\alpha\right)_\beta \phi_\beta\,. 
\end{equation}
By plugging the rotated generators from \eqref{eqn:rotatedgenh} into the expression for the mass matrix \eqref{eqn:massmatrix}, one finally obtains the requested diagonal form
\begin{equation}
  \bar M_{\alpha\beta} := \diag ( \lambda_\alpha )\,.
\end{equation}

The first order variation of the scalar potential and its vev vanish due to effective theory's field equation
\begin{equation}
  \frac{\delta V}{\delta \phi_\alpha} = 0\,.
\end{equation}
Here a projection like in \eqref{eqn:eomgenRicci} is not necessary, because the $\phi_\alpha$'s already describe the physical degrees of freedom only. Thus $V$ is only governed by second order perturbations, which lead to
\begin{equation}\label{eqn:expansionV}
  V = 2 \lambda_\alpha \phi_\alpha^2 + \mathcal{O}(\phi^3)\,.
\end{equation}
When inserting the expression for the generalized metric \eqref{eqn:genmetricfluctuation} into the kinetic term for the generalized metric in \eqref{eqn:ddimeffaction}, one obtains 
\begin{equation}\label{eqn:expansionKin}
  D_\mu \mathcal{H}_{MN} D^\mu \mathcal{H}^{MN} = \sum\limits_\alpha 4
    \partial_\mu \phi_\alpha \partial^\mu \phi_\alpha + \text{interaction terms}\,.
\end{equation}
The interaction terms describe self-couplings among the scalars $\phi_\alpha$ and couplings between scalars and gauge bosons $a_{M\mu}$, which are fluctuation around the vev of $A_{M\mu}$. The quadratic part of the Lagrangian for the scalars $\phi_\alpha$ is obtained by plugging \eqref{eqn:expansionV} and \eqref{eqn:expansionKin} into the action \eqref{eqn:ddimeffaction} and reads
\begin{equation}
  \mathcal{L}_\phi = \frac{1}{2} \sum\limits_\alpha \left( \partial_\mu \phi_\alpha
    \partial^\mu \phi_\alpha - 4 \lambda_\alpha \phi_\alpha^2 \right)\,.
\end{equation}
It identifies $2 \sqrt{\lambda_\alpha} = m_\alpha$ as the mass of the scalar field $\phi_\alpha$. Thus the eigenvalues $\lambda_\alpha$ have to be positive or zero in order to avoid tachyons. 
So we see that the string theory which belongs to this background should give rise to $(D-d)^2$ scalars $\phi_\alpha$ with the masses $m_\alpha$. Furthermore there should be $2(D-d)$ vector bosons $a_{M\mu}$ which arise from the internal symmetry of the scalars.

\subsection{Solution of flux constrains in $(D-d)=3$ dimensions}\label{sec:solutionsconstr}
In section~\ref{sec:DFTbackgrounds} and \ref{sec:minkowski}, we have discussed various constraints on the covariant fluxes. Only when all these constraints hold, one is able to construct a consistent background. Now we want to look systematically for their solutions. We restrict our search to $(D-d)=3$-dimensional compact spaces. In this case the number of compact dimensions is large enough to find interesting, non-trivial solutions. On the other hand it is still so small that we are able to manage the search with an appropriate effort.

As shown in \eqref{eqn:embeddingtensor}, there is a direct link between the covariant flux $\mathcal{F}\indices{_I_J^K}$ and the embedding tensor of gauged supergravities. For $(D-d)=3$, the $X_I$ in \eqref{eqn:embeddingtensor} describe the O$(3,3)$ generators labelled by $I=1,\dots,6$. Group-theoretically, $(X_I)\indices{_J^K}$ lives in the tensor product
\begin{equation}\label{eqn:decomptensorprod}
  6 \otimes 15 = 6 \oplus \overline{10} \oplus 10 \oplus 64 \,.
\end{equation}
The first factor in this product is the vector representation of SO$(3,3)$ and the second is the adjoint representation of SO$(3,3)$. There is one linear constraint, namely that the covariant fluxes are totally antisymmetric ($\mathcal{F}_{IJK} = \mathcal{F}_{[IJK]}$). This implies that the irreps $6$ and $64$ of the general tensor product decomposition  \eqref{eqn:decomptensorprod} are absent. The remaining irreps $\overline{10} \oplus 10$ matches perfectly the number of independent components of $\mathcal{F}_{IJK}$, which is $6\cdot5\cdot4/ 3! = 20$ in $2(D-d)=6$ dimensions.

Following the reasoning in \cite{Dibitetto:2012rk}, one can express $\left(X_I\right)\indices{_J^K}$ also as irreps of SL$(4)$, which is isomorphic to SO$(3,3)$. In this case \eqref{eqn:decomptensorprod} does not change. To distinguish between the two different groups, one introduces fundamental SL$(4)$ indices $p,q,r = 1,\dots,4$. The generators $\left(X_I\right)\indices{_J^K}$ can also be written in terms of SL$(4)$ indices
\begin{equation}\label{eqn:decompfluxes} 
  \left(X_{mn}\right)\indices{_p^q} = \frac{1}{2} \delta^q_{[m} M_{n]p} - 
    \frac{1}{4} \varepsilon_{mnpr}\tilde M^{rq}\,,
\end{equation}
where $M_{np}$ and $\tilde M^{rq}$ are symmetric $4\times 4$ matrices and $\varepsilon$ denotes the Levi-Civita symbol. The matrices $M_{np}$ and $\tilde M^{rq}$ have $4\cdot 5/2=10$ independent components each and hence match exactly the remaining irreps $\overline{10}$ and $10$ in \eqref{eqn:decomptensorprod}. A double index, like $mn$ in $\left(X_{mn}\right)\indices{_p^q}$, labels the $6$ independent components of the SL$(4)$ irrep $6$. These $6=4\cdot 3/2$ different components are the entries of an antisymmetric $4\times 4$ matrix. They are lowered by
\begin{equation}
  X_{mn} = \frac{1}{2} \varepsilon_{mnpq} X^{pq}\,.
\end{equation}
At this point, it is important to keep in mind that the indices $n$, $p$ of $M_{np}$ and $r$, $q$ of $\tilde M^{rq}$ are still fundamental SL$(4)$ indices and not doubled ones. Finally we transform the fundamental SL$(4)$ indices $p$ and $q$ in $\left(X_{mn}\right)\indices{_p^q}$ to double indices $pq$ and $rs$ respectively by using the identity
\begin{equation}\label{eqn:doubleindices}
  \left(X_{mn}\right)\indices{_{pq}^{rs}} \ = \ 
    2 \left(X_{mn}\right)\indices{_{[p}^{[r}} \delta^{s]}_{q]}\,.
\end{equation}
The covariant fluxes in this representation using $6$ of SL$(4)$ indices, are linked to one with $6$ of SO$(3,3)$ indices, used throughout the paper, by the 't Hooft symbols $(G_I)^{mn}$. For $(D-d)=3$, they are defined as
\begin{align}
  \left( G^1 \right)^{mn} &= \begin{pmatrix}
    0 & -1 & 0 & 0 \\
    1 & 0 & 0 & 0 \\
    0 & 0 & 0 & 0 \\
    0 & 0 & 0 & 0
  \end{pmatrix} &
  \left( G^2 \right)^{mn} &= \begin{pmatrix}
    0 & 0 & -1 & 0 \\
    0 & 0 & 0 & 0 \\
    1 & 0 & 0 & 0 \\
    0 & 0 & 0 & 0
  \end{pmatrix} &
  \left( G^3 \right)^{mn} &= \begin{pmatrix}
    0 & 0 & 0 & -1 \\
    0 & 0 & 0 & 0 \\
    0 & 0 & 0 & 0 \\
    1 & 0 & 0 & 0
  \end{pmatrix} \nonumber \\
  \left( G_1 \right)^{mn} &= \begin{pmatrix}
    0 & 0 & 0 & 0 \\
    0 & 0 & 0 & 0 \\
    0 & 0 & 0 & -1 \\
    0 & 0 & 1 & 0
  \end{pmatrix} &
  \left( G_2 \right)^{mn} &= \begin{pmatrix}
    0 & 0 & 0 & 0 \\
    0 & 0 & 0 & -1 \\
    0 & 0 & 0 & 0 \\
    0 & 1 & 0 & 0
  \end{pmatrix} &
  \left( G_3 \right)^{mn} &= \begin{pmatrix}
    0 & 0 & 0 & 0 \\
    0 & 0 & 1 & 0 \\
    0 & -1 & 0 & 0 \\
    0 & 0 & 0 & 0
  \end{pmatrix}
\end{align}
and fulfill the identities
\begin{align}
  \left( G_I \right)_{mn} \left( G_J \right)^{mn} &= 2\eta_{IJ}\,, \\
  \left( G_I \right)_{mp} \left( G_J \right)^{pn} +
  \left( G_J \right)_{mp} \left( G_I \right)^{pn} &= -\delta^n_m \eta_{IJ}\,, \\
  \left( G_I \right)_{mp} \left( G_J \right)^{pq} \left( G_K \right)_{qr}
  \left( G_L \right)^{rs} \left( G_M \right)_{st} \left( G_N \right)^{tn}
  &= \delta^n_m \varepsilon_{IJKLMN}\,.
\end{align}
Finally, we can express the covariant fluxes as 
\begin{equation}\label{eqn:SL(4)toSO(3,3)}
  \mathcal{F}_{IJK} = \left( X_{mn} \right)\indices{_{pq}^{rs}}
    \left( G_I \right)^{mn} \left( G_J \right)^{pq} \left( G_K \right)_{rs}
    \,.
\end{equation}
To evaluate the condition \eqref{eqn:fluxeseom}, which arise from the effective theory's equations of motion, one also needs the covariant fluxes in flat indices
\begin{equation}\label{eqn:fluxesflatvacuum}
  \mathcal{F}_{ABC} = \bar{E}\indices{_A^I} \bar{E}\indices{_B^J} \bar{E}\indices{_A^K} \mathcal{F}_{IJK}\,.
\end{equation}
This equation is invariant under O$({D-d},{D-d})$ transformations of the vacuum's generalized vielbein and the covariant fluxes, like
\begin{equation}\label{eqn:O(D-d,D-d)trafofluxes}
  \bar E\indices{_A^I} \rightarrow \bar E\indices{_A^J} O\indices{_J^I}
    \quad\text{and}\quad
  \mathcal{F}_{IJK} \rightarrow \mathcal{F}_{LMN} O\indices{^L_I} O\indices{^M_J} O\indices{^N_K}
    \quad\text{with}\quad
  O\indices{^M_N}\eta_{ML}O\indices{^L_K}=\eta_{NK} \,.
\end{equation}
Furthermore \eqref{eqn:fluxeseom} is invariant under double Lorentz transformations
\begin{equation}
  \bar E\indices{_A^I}\rightarrow T\indices{_A^B}\bar E\indices{_B^I}
    \quad \text{with} \quad
  T\indices{_A^C} \delta_{CD} T\indices{_B^D} = \delta_{AB}
    \quad \text{and} \quad
  T\indices{_A^C} \eta_{CD} T\indices{_B^D} = \eta_{AB}\,.
\end{equation}
Combining these two transformations, one is able to choose an arbitrary vacuum vielbein $\bar E\indices{_A^I}$. In the following, we use
\begin{equation}\label{eqn:defvacuumvielbein}
  \bar{E}\indices{_A^I} := \delta_A^I\;, 
\end{equation}
which allows to identify the components of the covariant fluxes in flat and curved indices. Other choices would be possible too, but they would make explicit calculations more complicated. This shows nicely that all relevant informations about the vacuum can be embedded in the covariant fluxes.

Next, we state and solve the constraints on the fluxes in terms of \eqref{eqn:SL(4)toSO(3,3)}. First, using the decomposition \eqref{eqn:decompfluxes}, the Jacobi-type constraint \eqref{eqn:quadraticc} on the fluxes reads
\begin{equation}\label{eqn:quadraticcMMtilde}
  M_{mp} \tilde M^{pn} = \frac{1}{4} \delta^m_n M_{qp} \tilde M^{pq}\,.
\end{equation}
Because $M_{np}$ is symmetric, it can always be diagonalized by an SO$(4)$ transformation. The group SO$(4)$ is the maximal compact subgroup of SL$(4)$ and it is, up to a discrete Z${}_2$, isomorphic to SO$(3)\times$SO$(3)$, the maximal compact subgroup of SO$(3,3)$. Hence it is always possible to diagonalize $M_{np}$ by an O$(3)\times$O$(3)$ double Lorentz transformation applied on the covariant fluxes. Such transformations leave all constraints on the covariant fluxes invariant. When $M_{np}$ is diagonal, $\tilde M_{rq}$ has to be diagonal, too. Otherwise the constraint \eqref{eqn:quadraticcMMtilde} is violated. In this case one can identify the components
\begin{equation}
  M_{mn} = \diag \begin{pmatrix} H_{123} & Q^{23}_1 & Q^{31}_2 
    & Q^{12}_3 \end{pmatrix} \quad \text{and} \quad
  \tilde{M}_{mn} = \diag \begin{pmatrix} R^{123} & f^1_{23} & f^2_{31} 
    & f^3_{12} \end{pmatrix}
\end{equation}
by applying \eqref{eqn:decompfluxes}, \eqref{eqn:doubleindices}, \eqref{eqn:SL(4)toSO(3,3)} and the mapping between the covariant fluxes $\mathcal{F}_{ABC}$ in flat indices and the $H$-, $f$-, $Q$- and $R$-flux derived in section~\ref{sec:covariantfluxes} successively. These remaining fluxes automatically fulfill
\begin{equation}
  \mathcal{F}\indices{^M_M_N} = 0 \quad \leftrightarrow \quad
    f^i_{ij} = 0 \quad \quad \text{and} \quad Q^{ij}_i = 0\,,
\end{equation}
as required by \eqref{eqn:FLLN=0}. 
\begin{table}[b]
 \centering
  \begin{tabular}{|c||c|c|c|c|}
    \hline
    $\alpha$ & $m_\alpha $ & 
      $( \bar h_\alpha )_{ij}$ & 
      $( \bar h_\alpha )\indices{^k_j}$ & $\bar\phi_\alpha$ \\
    \hline
    1 & $2 \left|f\right|$ &
      $\begin{pmatrix}
        0 & 0 & 0 \\
        0 & -1 & 0 \\
        0 & 0 & 1 
      \end{pmatrix}$ & $0$ & $\tau_\compliI$ \\
    \hline
    2 & $2 \left|f\right|$ &
      $\begin{pmatrix}
        0 & 0 & 0 \\
        0 & 0 & 1 \\
        0 & 1 & 0 
      \end{pmatrix}$ & $0$ & $\tau_\compliR$ \\
    \hline
    3 & $2 \left|H\right|$ &
      $\begin{pmatrix}
        0 & 0 & 0 \\
        0 & 1 & 0 \\
        0 & 0 & 1 
      \end{pmatrix}$ & $0$ & $\rho_\compliI$ \\
    \hline
    4 & $2 \left|H\right|$ & $0$ & 
      $\begin{pmatrix}
        0 & 0 & 0 \\
        0 & 0 & -1 \\
        0 & 1 & 0 
      \end{pmatrix}$ & $\rho_\compliR$ \\
    \hline
  \end{tabular}
  \caption{The massive scalar fields with arise from the fluxes in \eqref{eqn:solutionfluxes}.}\label{tab:massivescalars}
\end{table}
Hence, according to \eqref{eqn:strongconstfluxes}, the strong constraint restricts the fluxes by
\begin{equation}
  H_{123} R^{123} + Q^{23}_1 f^1_{23} + Q^{31}_2 f^2_{31} + 
    Q^{12}_3 f^3_{12} = M_{qp} {\tilde M}^{pq} = 0\,.
\end{equation}
In conjunction with the quadratic constraint \eqref{eqn:quadraticcMMtilde} 
this gives rise to
\begin{equation}
  H_{123} R^{123} = 0\,, \qquad  
  Q^{23}_1 f^1_{23} = 0\,, \qquad
  Q^{31}_2 f^2_{31} = 0\,, \qquad 
  Q^{12}_3 f^3_{12} = 0\,.
\end{equation}
Finally, the constraint from the field equations \eqref{eqn:fluxeseom} leads to
\begin{align}
  \left( H_{123} - Q^{23}_1 \right)^2 - 
    \left( Q^{31}_2 - Q^{12}_3 \right)^2 &=
    \left( R^{123} - f^1_{23} \right)^2 -
    \left( f^2_{31} - f^3_{12} \right)^2 
    \\
  \left( H_{123} - Q^{31}_2 \right)^2 -
    \left( Q^{12}_3 -  Q^{23}_1 \right)^2 &=
    \left( R^{123} - f^2_{31} \right)^2 -
    \left( f^3_{12} - f^1_{23} \right)^2 \\
  \left( H_{123} - Q^{12}_3 \right)^2 -
    \left( Q^{23}_1 - Q^{31}_2 \right)^2 &=
    \left( R^{123} - f^3_{12} \right)^2 -
    \left( f^1_{23} - f^2_{31} \right)^2 \,.
\end{align}
The only non-trivial solution for these three equations, which is not excluded by the strong constraint, is
\begin{equation}\label{eqn:solutionfluxes}
  H_{123} = Q^{23}_1 = H\,, \quad  Q^{31}_2 = Q^{12}_3 = 0\,, \quad
  R^{123} = f^1_{23} = 0  \quad \text{and} \quad f^2_{31} = f^3_{12} = f\,.
\end{equation}
In $D-d=3$ dimensions, only these fluxes are allowed for backgrounds without a warp factor. This shows how restrictive the conditions on the covariant fluxes are. The covariant fluxes in \eqref{eqn:solutionfluxes} are given in flat indices. Thus they are invariant under $O({D-d},{D-d})$ transformations \eqref{eqn:O(D-d,D-d)trafofluxes} but depend on the fixing of the double Lorentz symmetry. In total, we obtain three different kind of solutions which will be discussed in section~\ref{sec:twistforD-d=3} in detail:

\begin{itemize}
  \item $f\neq0$, $H=0$: this is a geometric background, called {\it single elliptic} $f$-flux space.
  \item $f=0$, $H\neq0$: this is a non-geometric background, because by \eqref{eqn:solutionfluxes} it has non-vanishing 
  $H$ \textit{and} $Q$ flux. It is 
  called {\it single elliptic} $H,Q$-flux space. It is, however, T-dual to the previous, geometric background.
  \item $f\neq0$, $H\neq0$: this is a non-geometric background, called {\it double elliptic} $f,H,Q$-flux space. It is not T-dual to any geometric space.
\end{itemize}

Following the reasoning in section~\ref{sec:spectrum} one is able to express the fluctuations of the generalized metric around its vev as
\begin{equation}
  \delta \mathcal{H}^{MN} = \sum\limits_\alpha \bar{E}\indices{_A^M} \bar{E}\indices{_B^N} (\bar h_\alpha)^{AB} \phi_\alpha \,.
\end{equation}
By using $\bar{E}\indices{_A^M}=\delta_A^M$, c.f.~\eqref{eqn:defvacuumvielbein}, it is straightforward to identify such fluctuation of the generalized metric \eqref{eqn:genmetricBg} with
\begin{equation}
  \delta g_{ij} = \sum\limits_\alpha ( \bar h_\alpha )_{ij} \phi_\alpha
  \quad \text{and} \quad 
  \delta B_{ij} = \delta_{ik} ( \bar h_\alpha )\indices{^k_j} \,.
\end{equation}
For the double elliptic background, there are in total four massive and five massless scalar fields. The massive ones are listed in table~\ref{tab:massivescalars}. In the directions $y^2$ and $y^3$ the shape of the double tours specified by $\bar{\mathcal H}^{MN}$ is completely fixed by the massive scalars. A double torus in these directions is parameterized by four real scalars which correspond the metric components $g_{22}$, $g_{33}$, $g_{23}$ and the $B$-field component $B_{23}$. They can also be expressed in terms of the complex structure $\tau=\tau_\compliR+i\tau_\compliI$ and the K\"ahler parameter  $\rho=\rho_\compliR + i\rho_\compliI$ as
\begin{equation}\label{eqn:metricbrhotau}
  \begin{pmatrix}
    g_{22} & g_{23} \\
    g_{23} & g_{33}
  \end{pmatrix} =
  \frac{\rho_\compliR}{\tau_\compliI} \begin{pmatrix}
    1 & \tau_\compliR \\
    \tau_\compliR & \left| \tau_\compliR \right|^2
  \end{pmatrix} \quad \text{and} \quad
  - B_{23} = B_{32} = \rho_\compliR\,. 
\end{equation}
For $\bar{\mathcal H}^{MN}=\delta^{MN}$, one gets $\bar \tau_\compliI = \bar \rho_\compliI = 1$ and $\bar \tau_\compliR = \bar \rho_\compliR = 0$. Here the bar on $\tau$, $\rho$ and its component $\tau_\compliR$, $\tau_\compliI$, $\rho_\compliR$ and $\rho_\compliI$ does not indicates complex conjugation, but that these quantities belong to the vacuum vielbein $\bar E\indices{_A^M}$. The variation of the metric and the $B$-field in \eqref{eqn:metricbrhotau} with respect to $\tau_\compliR$, $\rho_\compliR$, $\tau_\compliI$ and $\rho_\compliI$ leads to the same results as given in table~\ref{tab:massivescalars}. 
Hence it is straightforward to identify the scalar moduli $\phi_\alpha$ in this table with the real and imaginary parts of $\tau$ and $\rho$. The full scalar potential in these moduli reads
\begin{equation}
  \label{eqn:explicitspotential}
  V = \frac{f^2 \left(1 + 2(\tau_\compliR^2 - \tau_\compliI^2) + |\tau|^4\right)}{2 \tau_\compliI^2}
   + \frac{H^2 \left(1 + 2(\rho_\compliR^2 - \rho_\compliI^2) + |\rho|^4\right)}{2 \rho_\compliI^2} \,.
\end{equation}
A minimum of this potential has to fulfill
\begin{equation}
  \left.\frac{\partial V}{\partial \tau_\compliR}\right|_{\tau=\bar \tau} = \frac{f^2 \bar \tau_\compliR(1+|\bar \tau|^2)}{\bar \tau_\compliI^2} = 0 \quad\text{and}\quad
  \left.\frac{\partial V}{\partial\tau_\compliI}\right|_{\tau=\bar \tau} = \frac{f^2 \left[ 2 \bar \tau_\compliR^2 (\bar \tau_\compliI^2 - 1) + 2\bar \tau_\compliI^4 - |\bar \tau|^4 - 1\right]}{\bar \tau_\compliI^3}=0\,.
\end{equation}
From the first equation follows that $\bar \tau_\compliR=0$. In this case, the second one simplifies to $\bar \tau_\compliI^4=1$ and thus gives rise to $\bar \tau_\compliI=1$. These are exactly the values we expected. The same argumentation holds for $\rho$. Plugging the vevs $\bar \tau$ and $\bar \rho$ into \eqref{eqn:explicitspotential}, we see that the scalar potential $V(\bar \tau,\bar \rho)=0$ vanishes for the vacuum. This result is in accordance with \eqref{eqn:minkowskivacuum}. After a short calculation, one obtains the Hesse matrix
\begin{equation}
  \left.\frac{\partial^2 V}{\partial \phi_\alpha  \partial \phi_\beta}\right|_{\bar \phi} = 4 \begin{pmatrix}
    f^2 & 0 & 0 & 0 \\
    0 & f^2 & 0 & 0 \\
    0 & 0 & H^2 & 0 \\
    0 & 0 & 0 & H^2
  \end{pmatrix}
    \quad\text{with}\quad
  \phi = \{\tau_\compliR, \tau_\compliI, \rho_\compliR, \rho_\compliI\}
\end{equation}
for the vacuum. It is diagonal and so proves that $\tau$ and $\rho$ are indeed the right moduli to describe the massive scalar field which arise in the effective theory.

\section{Twists, Killing vectors and background fields}\label{sec:constrtwists}
Until now, we have only considered the constant values of the covariant fluxes $\mathcal{F}_{IJK}$. But in order to construct the metric and $B$-field or $\beta$-field of a doubled geometry, one needs to know the twist $U\indices{^M_N}$ and its action on the scalar fields $\widehat{\cal H}^{MN}$. Here we give twists that reproduce the given covariant fluxes. We focus on covariant fluxes that describe fibered backgrounds. For them, we are able to provide an explicit expression for the twist and also for the Killing vectors which are associated to it. The background described in section~\ref{eqn:solutionfluxes} is such a fibration. Hence we can apply these results to study its properties in more detail. Finally we show how the remaining double Lorentz symmetry of the covariant fluxes is fixed, for which there are different possibilities related to each other via a field redefinitions.

\subsection{Fibered backgrounds}
To construct explicit expressions for the twist $U\indices{^M_N}$ and its Killing vectors, we focus on fibered geometries ${\cal M}^{2(D-d)}$ of the kind
\begin{equation}
  T^{2 d_\mathrm{f}}\,\hookrightarrow\, {\cal M}^{2(D-d)}
    \,\hookrightarrow\,T^{2 d_\mathrm{b}}\,.
\end{equation}
Here $T^{2 d_\mathrm{f}}$ is a $2 d_\mathrm{f}$-dimensional double torus in the fiber, which is twisted by the covariant fluxes. While the $2d_b$-dimensional, rectangular base torus $T^{2d_\mathrm{b}}$ is not affected by this twist. At first glance this sounds like a strong limitation, which excludes many potential backgrounds. Nevertheless, the consistent backgrounds from section~\ref{sec:solutionsconstr}, which satisfy the various constraints discussed in this paper, are exactly of this form. In order to make the structure of the fibration manifest, we split the $2(D-d)$ internal, compact coordinates $Y^{\mathcal M}=\begin{pmatrix} \tilde{y}_i & y^i \end{pmatrix}$ into
\begin{equation}
  Y^{\hat M} = \begin{pmatrix} Y^{\tilde M} & Y^M \end{pmatrix}\,.
\end{equation}
Indices with a tilde label the base coordinates and indices without a tilde are assigned to the directions of the fiber. For these conventions, the invariant metric is given by
\begin{equation}
  \eta^{\hat M \hat N} = \begin{pmatrix} \eta^{\tilde M\tilde N} & 0 \\
    0 & \eta^{MN}
  \end{pmatrix}\,.
\end{equation}
Analogous expressions hold for the generalized vielbein, the twist and the parameter of generalized diffeomorphisms.
Using this splitting, the twist $U\indices{_{\hat N}^{\hat M}}$ can be expressed by the matrix exponential
\begin{equation}\label{eqn:twistfromflux}
  U\indices{_{\hat N}^{\hat M}}(Y^{\tilde I}) = \exp\left( \mathcal{F}
    \indices{_{\hat N}^{\hat M}_{\tilde I}} Y^{\tilde I} \right) \,.
\end{equation}
The only non-vanishing covariant fluxes are $\mathcal{F}_{NM\tilde I}$, while the remaining flux components 
\begin{equation}\label{eqn:compfluxes=0}
  \mathcal{F}_{\hat N\hat M I}=0  \quad \text{and} \quad 
  \mathcal{F}_{\tilde N \tilde M \hat I}=0
\end{equation}
vanish in order to be compatible with the fibration discussed above. Furthermore, we consider only matrices in the exponent of \eqref{eqn:twistfromflux}, which commute for arbitrary values of $\tilde I$ and $\tilde J$. Thus the additional constraint
\begin{equation}\label{eqn:commutatorconstr}
  \mathcal{F}\indices{_{\tilde I}^M_L} \mathcal{F}\indices{_{\tilde J}^L_N} -
  \mathcal{F}\indices{_{\tilde J}^M_L} \mathcal{F}\indices{_{\tilde I}^L_N} 
  = 0 \quad \text{or} \quad
  \mathcal{F}_{LM[\tilde I} \mathcal{F}\indices{^L_{\tilde J ]}_N} = 0
\end{equation}
has to hold. Without it and \eqref{eqn:compfluxes=0}, we are not able to derive the following properties of the twist:
\begin{equation}\label{eqn:propUfibration}
  U\indices{_{\hat N}^{\tilde M}} = \delta_{\hat N}^{\tilde M} \,, \quad 
  U\indices{_{\tilde N}^{\hat M}} = \delta_{\tilde N}^{\hat N} \quad \text{and} \quad
  \partial_{\hat L} U\indices{_{\hat N}^{\hat M}} = \begin{cases} 
    \mathcal{F}\indices{_N^P_{\tilde L}} U\indices{_P^M} & \\
    0 & \text{otherwise.}
  \end{cases}
\end{equation}
With them, it is then straightforward to calculate the non-vanishing coefficients of anholonomy
\begin{equation}\label{eqn:FfromU}
  \Omega_{\tilde I J K} = \partial_{\tilde I} U\indices{_J^M} 
    U_{KM} = \mathcal{F}\indices{_J^N_{\tilde I}} U\indices{_N^M} U_{KM} = \mathcal{F}\indices{_J^N_{\tilde
      I}} \eta_{NK} =  \mathcal{F}_{\tilde I J K}.
\end{equation}
The remaining components  
\begin{equation}
  \Omega_{I \tilde J K} = -\Omega_{I K \tilde J} = 0
\end{equation}
vanish. Hence, the non-vanishing components of the covariant fluxes for the twist \eqref{eqn:twistfromflux} are
\begin{equation}
  \mathcal{F}_{\tilde I J K} = \Omega_{\tilde I J K} +
    \Omega_{K \tilde I J} + \Omega_{J K \tilde I} = \Omega_{\tilde I J K}\;, 
\end{equation}
as expected.

Furthermore we have to find the $2(D-d)$ Killing vectors $K\indices{_{\hat I}^{\hat J}}$ connected to the twist $U\indices{_{\hat N}^{\hat M}}$. For the fibration, discussed in this section, they are given by
\begin{equation}
  K\indices{_{\hat I}^{\hat J}} = \exp\left( -\frac{1}{2} \mathcal F\indices{_{\tilde I}^{\hat J}_{\hat L}} Y^{\hat L} \right) \,.
\end{equation}
Here the $\tilde I$ in $\mathcal{F}\indices{_{\tilde I}^{\hat J}_{\hat L}} Y^{\hat L}$ denotes that the matrix given by this expression has only non-vanishing entries in columns with are associated to base coordinates. Again, we find the following properties:
\begin{equation}\label{eqn:Killingfromfluxes}
  K\indices{_{\hat I}^{\tilde J}}=\delta_{\hat I}^{\tilde J}\,, \quad
  K\indices{_I^{\hat J}} = \delta_I^{\hat J} \quad \text{and} \quad
  \partial_{\hat L} K\indices{_{\hat I}^{\hat J}} = \begin{cases}
    - \frac{1}{2} \mathcal{F}\indices{_{\tilde I}^J_L} & \\
    0 & \text{otherwise.}
  \end{cases}
\end{equation}
With these identities, it is straightforward to show that
\begin{align}
  \mathcal{L}_{K\indices{_{\hat I}^{\hat J}}} U\indices{_{\hat N}^{\hat M}} &= K\indices{_{\hat I}^{\tilde P}} \partial_{\tilde P} U\indices{_N^M} + \partial^M K\indices{_{\tilde I}^{P}} U_{\hat N P} - U\indices{_{\hat N}^P} \partial_P K\indices{_{\tilde I}^M} \nonumber \\
  &= \mathcal{F}\indices{_N^P_{\tilde I}} U\indices{_P^M} - \frac{1}{2}\mathcal{F}\indices{_{\tilde I}^P^M} U_{NP} + \frac{1}{2} U\indices{_N^P} \mathcal{F}\indices{_{\tilde I}^M_P} \nonumber \\
  &= \mathcal{F}\indices{_{\tilde I}_N^P} U\indices{_P^M} - U\indices{_N^P} \mathcal{F}\indices{_{\tilde I}_P^M} = [\left(\mathcal{F}_{\tilde I}\right), U] = 0 \,.
\end{align}
In the last step we have used that according to \eqref{eqn:commutatorconstr} the matrices $\left(\mathcal{F}_{\tilde I}\right)\indices{_N^M}$ have to commute for all possible values of $\tilde I$. We can check that the condition
\begin{equation}
  \partial_{\hat J} K\indices{_{\hat I}^{\hat J}} = -\frac{1}{2} \mathcal{F}\indices{_{\tilde I}^J_J} = 0 \quad \leftrightarrow \quad \mathcal{F}\indices{^{\hat L}_{\hat L}_{\hat N}} = 0
\end{equation}
holds. According to \eqref{eqn:invdensity} it has to be fulfilled in order to leave densities invariant when they are shifted along the Killing vectors. For the fibrations discussed here, this condition is equivalent to \eqref{eqn:FLLN=0}. Finally, we calculate the structure coefficients associated to the algebra generated
 by the Killing vectors. According to \eqref{eqn:structurekilling}, they read
\begin{equation}\label{eqn:structurecoeffkillingvecvstwist}
  \tilde{\mathcal F}\indices{_{\hat I}_{\hat J}_{\hat K}} = -\frac{1}{2} \mathcal{F}\indices{_{\hat I}_{\hat J}_{\hat K}}\,.
\end{equation}
Despite having the same structure coefficients up to a factor -1/2, the Killing vectors have very different properties in comparison to the twist. In general, $K\indices{_{\hat I}^{\hat J}}$ is not an O$({D-d},{D-d})$ valued matrix. Furthermore, if $U\indices{_{\hat N}^{\hat M}}$ fulfills the strong constraint, it is not guaranteed that the Killing vectors also do so. Nevertheless, the construction in this section guarantees that their algebra is closed.

The value of the twist after going completely around the base circle in the direction $\tilde I$ is called monodromy. It is given by the expression
\begin{equation}
  M\indices{_{\tilde I}_N^M}  = \exp \left( 2\pi   
    \mathcal{F}\indices{_{\tilde I}_N^M} \right)
\end{equation}
and has to be O$({D-d},{D-d},\mathds{Z})$ valued. When only considering pure DFT, an O$({D-d},{D-d})$ valued monodromy would be sufficient. In this case the two different tori at $Y^{\tilde I}=0$ and $Y^{\tilde I} = 2\pi$ can be identified by a generalized diffeomorphism. But in string theory tori are only identified by the subgroup O$({D-d},{D-d},\mathds{Z})$ whose elements parameterize T-duality transformations. As we will show in the following section, this restriction allows only for discrete values for covariant fluxes.

\subsection{Configurations with Minkowski vacuum}\label{sec:twistforD-d=3}
Section~\ref{sec:solutionsconstr} has already presented covariant fluxes, which fulfill the various constraints imposed in section~\ref{sec:DFTbackgrounds} and lead to a Minkowski vacuum in the external directions. Additionally, these fluxes satisfy \eqref{eqn:commutatorconstr} and give rise to a fibered background with $d_\mathrm{f}=2$ and $d_\mathrm{b}=1$. Thus we are able to construct the associated twist $U\indices{^M_N}$ and the Killing vectors $K\indices{_I^J}$.

For $d_\mathrm{f}=2$, the twist of the fiber is an element of O$(2,2)$. Such an element can be decomposed into SO$(2,2) \times \mathrm{Z}_2$. The $\mathrm{Z}_2$ part consists of two elements, the identity and an O$(2,2)$ element T with $\det T=-1$ and $T^2 = 1$. Here we choose $T$ as a T-duality transformation along the second direction of the fiber, which amounts to
\begin{equation}
 T=\begin{pmatrix}
    1 & 0 & 0 & 0 \\
    0 & 0 & 0 & 1 \\
    0 & 0 & 1 & 0 \\
    0 & 1 & 0 & 0
  \end{pmatrix}=T^{-1}=T^T\,.
\end{equation}
The SO$(2,2)$ part decomposes into $\mathrm{SL}(2)_\tau\times \mathrm{SL}(2)_\rho$. Thus, in order to express an SO$(2,2)$ element, one needs two SL$(2)$ matrices, which we call $M_\tau$ and $M_\rho$. They are mapped to the corresponding SO$(2,2)$ element $M$ by
\begin{equation}\label{eqn:decompSL2xSL2}
  M = \begin{pmatrix}
    M_\tau & 0 \\
    0 & M_\tau^{-T}
  \end{pmatrix}
  T \begin{pmatrix}
    M_\rho & 0 \\
    0 & M_\rho^{-T}
  \end{pmatrix} T^{-1}\,.
\end{equation}
We interpret $\tau$ as the complex structure and $\rho$ as the K\"ahler parameter of a torus in the fiber. $SL(2)$ transformations act on these two parameters as 
\begin{equation}\label{eqn:actiontau&rho}
  \tau' = \frac{a \tau + b}{c \tau + d} \quad \leftrightarrow \quad M_\tau =
    \begin{pmatrix} a & b \\ c & d \end{pmatrix}
  \quad \text{and} \quad
  \rho' = \frac{a' \rho +  b'}{c' \rho + d'} \quad \leftrightarrow \quad M_\rho =
    \begin{pmatrix} a' & b' \\ c' & d' \end{pmatrix}\;, 
\end{equation}
respectively. The T-duality transformation $T$ acts as an exchange of $\tau$ and $\rho$. More precisely, 
the isomorphism reads 
\begin{equation}\label{eqn:decompO(2,2)}
 \mathrm{O}(2,2) \cong \mathrm{SL}_\tau(2) \times \mathrm{SL}_\rho(2) 
  \times \mathrm{Z}_2^{\tau\leftrightarrow\rho}\;. 
\end{equation}
A convenient way to characterize SL$(2)$ group elements is given by their conjugacy classes. In total there are three different classes, which are discriminated by the traces
\begin{equation}
  \left| \Tr M \right| < 2 \quad \text{elliptic} \qquad 
  \left| \Tr M \right| = 2 \quad \text{parabolic} \qquad \text{and} \qquad
  \left| \Tr M \right| > 2 \quad \text{hyperbolic}
\end{equation}
of the corresponding SL$(2)$ element $M$.

By explicitly evaluating \eqref{eqn:twistfromflux} with the covariant fluxes obtained in \eqref{eqn:solutionfluxes}, 
we obtain the twist
\begin{equation}\label{eqn:twistHf}
  U\indices{^{\hat M}_{\hat N}}(x^1) = \begin{pmatrix}
    1 & 0 & 0 & 0 & 0 & 0 \\
    0 & 1 & 0 & 0 & 0 & 0 \\
    0 & 0 & \cos f x^1 \cos H x^1 & \sin f x^1 \cos H x^1  &
      - \sin f x^1 \sin H x^1 & \cos f x^1 \sin H x^1 \\
    0 & 0 & - \sin f x^1 \cos H x^1 & \cos f x^1 \cos H x^1 &
      - \cos f x^1 \sin H x^1 & - \sin f x^1 \sin H x^1 \\
    0 & 0 & - \sin f x^1 \sin H x^1 & \cos f x^1 \sin H x^1 &
      \cos f x^1 \cos H x^1 & \sin f x^1 \cos H x^1 \\
    0 & 0 & - \cos f x^1 \sin H x^1 & -\sin f x^1 \sin H x^1 &
      - \sin f x^1 \cos H x^1 & - \cos f x^1 \cos H x^1
  \end{pmatrix}
\end{equation}
and with \eqref{eqn:decompSL2xSL2} we are able to decompose this result into 
\begin{equation}\label{eqn:Utau&rho}
  U_\tau(x^1) = \begin{pmatrix}
    \cos f x^1 & \sin f x^1 \\
   -\sin f x^1 & \cos f x^1
  \end{pmatrix} \quad \text{and} \quad
  U_\rho(x^1) = \begin{pmatrix}
    \cos H x^1 & \sin H x^1 \\
   -\sin H x^1 & \cos H x^1
  \end{pmatrix}\,.
\end{equation}
These twist $U_\tau$ and $U_\rho$ are both elliptic. Each of them is an element of SO$(2)$, the maximal compact subgroup of SL$(2)$. As already stated, the possible values of $H$ and $f$  are not continuous because the monodromy
\begin{equation}
  M\indices{_1^M_N} = U\indices{^M_N}(2\pi)
\end{equation}
has to be an element of O$(2,2,\mathds{Z})$. This subset of O$(2,2)$
decomposes along the lines of \eqref{eqn:decompO(2,2)} into
\begin{equation}
  \mathrm{O}(2,2,\mathds{Z})\cong\mathrm{SL}(2,\mathds{Z})_\tau\times 
  \mathrm{SL}(2,\mathds{Z})_\rho\times \mathrm{Z}_2^{\tau\leftrightarrow
    \rho}\,.
\end{equation}
The discrete transformation is not realized by the monodromy. But the remaining two $SL(2,Z)$ transformations are not trivial and lead to
\begin{equation}
  M_\tau = \begin{pmatrix}
    \cos 2\pi f & \sin 2\pi f \\
   -\sin 2\pi f & \cos 2\pi f
  \end{pmatrix} \quad \text{and} \quad
  M_\rho = \begin{pmatrix}
    \cos 2\pi H & \sin 2\pi H \\
   -\sin 2\pi H & \cos 2\pi H
  \end{pmatrix}\,.
\end{equation}
\begin{table}[b]
  \centering
  \begin{tabular}{|c||c|c|}
    \hline
    $f \bmod 1$ & $\Tr M_\tau$ & $\bar \tau$ \\
    \hline
    $0$ & $2$  & $i$ \\
    $1/6$ & $1$  & $(-1+\sqrt{3} i)/2$ \\
    $1/4$ & $0$  & $i$ \\
    $1/3$ & $-1$ & $(-1+\sqrt{3} i)/2$ \\
    $1/2$ & $-2$ & $i$ \\
    \hline
  \end{tabular}\qquad\qquad\qquad
  \begin{tabular}{|c||c|c|}
    \hline
    $H \bmod 1$ & $\Tr M_\rho$ & $\bar \rho$ \\
    \hline
    $0$ & $2$  & $i$ \\
    $1/6$ & $1$  & $(-1+\sqrt{3} i)/2$ \\
    $1/4$ & $0$  & $i$ \\
    $1/3$ & $-1$ & $(-1+\sqrt{3} i)/2$ \\
    $1/2$ & $-2$ & $i$ \\
    \hline
  \end{tabular}
  \caption{Quantized values for the fluxes $f$ and $h$ and the corresponding vevs for $\tau$ and $\rho$.}\label{tab:quantizedfluxes}
\end{table}%
Each of these two matrices have to be an element of SL$(2,\mathds{Z})$, which is obviously the case if $f\mod 1$ and $H\mod 1$ are elements of the set $0$, $1/2$ or $1/4$. But this is not an exhaustive list of all allowed fluxes. We can still apply an $O(2,2)$ transformation \eqref{eqn:O(D-d,D-d)trafofluxes} to make the monodromies $M_\rho$ and $M_\tau$ elements of SL$(2,\mathds{Z})$. This is possible when both of them have integer traces. Table~\ref{tab:quantizedfluxes} lists all different values for the fluxes which fulfill this constraint. According to \eqref{eqn:O(D-d,D-d)trafofluxes} the vacuum vielbein $\bar E\indices{_A^M}$ gets modified by such transformations, too. Thus, the table also lists the new vevs for $\tau$ and $\rho$, respectively. The covariant fluxes in flat indices $\mathcal{F}_{ABC}$ are not affected by \eqref{eqn:O(D-d,D-d)trafofluxes} and their curved counterparts $\mathcal{F}_{IJK}$ are calculated from them with the vacuum vielbein $\bar E\indices{_A^M}(\bar\tau,\bar\rho)$ according to \eqref{eqn:fluxesflatvacuum}. Finally, a transformation into barred indices gives some additional insights into the structure of the monodromy
\begin{align}
  M\indices{^{\bar M}_{\bar N}} &= 
    R\indices{^{\bar M}_L} M\indices{^L_K} R\indices{^K_{\bar N}} \nonumber\\ &=
    \begin{pmatrix}
      \cos\left[ 2\pi(f - H) \right] & \sin\left[ 2\pi(f - H) \right] & 0 & 0 \\
      -\sin\left[ 2\pi(f - H) \right] & \cos\left[ 2\pi(f - H) \right] & 0 & 0 \\
      0 & 0 & \cos\left[ 2\pi(f + H) \right] & \sin\left[ 2\pi(f + H) \right] \\
      0 & 0 & -\sin\left[ 2\pi(f + H) \right] & \cos\left[ 2\pi(f + H) \right] 
    \end{pmatrix}\,.
\end{align}
Remembering that the first two rows describe the string's right moving part and the remaining ones the left moving part, it is obvious that this background is totally symmetric for $H=0, f\ne 0$ and totally asymmetric for $H\ne0, f=0$. 

According to \eqref{eqn:Killingfromfluxes}, the Killing vectors read
\begin{equation}
  K\indices{_{\hat I}^{\hat J}} = \begin{pmatrix}
    1 & 0 & 0 & 0 & 0 & 0 \\
    0 & 1 & -\frac{1}{2}( H x^3 + f \tilde x^3) & \frac{1}{2}( H x^2 + f \tilde x^2) & 
      -\frac{1}{2}( f x^3 + H \tilde x^3) & \frac{1}{2}( f x^2 + H \tilde x^2 ) \\
    0 & 0 & 1 & 0 & 0 & 0 \\
    0 & 0 & 0 & 1 & 0 & 0 \\
    0 & 0 & 0 & 0 & 1 & 0 \\
    0 & 0 & 0 & 0 & 0 & 1 \\
  \end{pmatrix} \,.
\end{equation}
They cannot be combined to an O$(3,3)$ valued matrix and for $H\ne 0$, $f\ne 0$, $K\indices{_2^{\hat J}}$ violates the strong constraint. Nevertheless the algebra of infinitesimal transformations along the Killing vectors closes. The only non-trivial Killing vector $K\indices{_2^{\hat J}}$ can be decomposed into
\begin{align}
  K\indices{_2^{\hat J}} &= K'^{\hat J} + K''^{\hat J} \quad \text{with} \\
  K'^{\hat J} &= \begin{pmatrix} 0 & \frac{1}{2} & -\frac{1}{2} H x^3 & \frac{1}{2} H x^2 & -\frac{1}{2} f x^3 & \frac{1}{2} f x^2 \end{pmatrix} \quad \text{and} \\
  K''^{\hat J} &= \begin{pmatrix} 0 & \frac{1}{2} & -\frac{1}{2} f \tilde x^3 & \frac{1}{2} f \tilde x^2 & -\frac{1}{2} H \tilde x^3 & \frac{1}{2} H \tilde x^2 \end{pmatrix}\,.
\end{align}
$K''^{\hat J}$ is equivalent to $K'^{\hat J}$ after a T-duality along all fiber directions. $K'^{\hat J}$ describes a coordinate transformation and a $B$-field gauge transformation, while its T-dual $K''^{\hat J}$ describes a coordinate transformation and a $\beta$-field gauge transformation. Thus, for $H\ne 0$ and $f\ne 0$, two coordinate patches of the background are always connected to each other by all possible kinds of generalized diffeomorphism: coordinate transformation, $B$- and $\beta$-field gauge transformation at the same time. This clearly shows that the double elliptic case cannot be discussed in SUGRA or even not in Generalized Geometry, because in these theories only two different kinds of generalized diffeomorphisms are allowed at the same time.

We close this section, by discussing a chain of T-dualities for the background specified by the twist \eqref{eqn:twistHf}. Such chains are well know from the torus with constant $H$-flux
\begin{equation}
  H_{ijk} \stackrel{T_{i}}{\longrightarrow} f^{i}_{jk}
    \stackrel{T_{j}}{\longrightarrow} Q_{k}^{ij}
    \stackrel{T_{k}}{\longrightarrow} R^{ijk}\,.
\end{equation}
A T-duality transformation along the $i$-th direction is given in terms of the $O({D-d},{D-d})$ element
\begin{equation}
  O\indices{^M_N} = \begin{pmatrix}
    \mathds{1} - m_i & m_i \\
    m_i & \mathds{1} - m_i
  \end{pmatrix}\,,
\end{equation}
where $m_i$ is a diagonal matrix with a one in the direction $i$, on which 
T-duality is performed and zeros in the other directions. In contrast to \eqref{eqn:O(D-d,D-d)trafofluxes}, T-duality act on the covariant fluxes only. It does not change the vaccum vielbein $\widehat{\bar E}\indices{^A_M}$. Hence, the covariant fluxes $\mathcal{F}_{IJK}$ transform like any other covariant object under T-duality, namely as
\begin{equation}
  {\mathcal F}'_{IJK} = \mathcal{F}_{LMN} O\indices{^L_I} O\indices{^M_J}    O\indices{^N_K}\,.
\end{equation}
When we start with the fluxes in \eqref{eqn:solutionfluxes} and do successively T-duality transformations along $x^2$, $x^3$ (isometric directions) and finally also over $x^1$, we obtain the T-dual configurations listed in table~\ref{tab:tdualitychain}.
\begin{table}[t]
  \centering
  \begin{tabular}{|c|c|c||c|ccc|ccc|c|}
    \hline
    $x^1$ & $x^2$ & $x^3$ & $H_{123}$ & $f^1_{23}$ & $f^2_{31}$ & 
      $f^3_{12}$ & $Q_1^{23}$ & $Q_2^{31}$ & $Q_3^{12}$ & $R^{123}$ \\
    \hline
    & & & $H$ & 0 & $f$ & $f$ & $H$ & 0 & 0 & 0 \\
    & $\bullet$ & & $f$ & 0 & $H$ & $H$ & $f$ & 0 & 0 & 0 \\
    & $\bullet$ & $\bullet$ & $H$ & 0 & $f$ & $f$ & $H$ & 0 & 0 & 0 \\
    $\bullet$ & $\bullet$ & $\bullet$ & 0 & $H$ & 0 & 0 & 0 &
      $f$ & $f$ & $H$ \\
    \hline
  \end{tabular}
  \caption{T-duality chain for the double elliptic background. Directions, on which T-duality was applied, are marked by a dot.}\label{tab:tdualitychain}
\end{table}%
Here, let us distinguish between the three different cases:
\begin{itemize}
  \item {\it Single elliptic space} with $f\neq0$, $H=0$: It is a geometric space with geometric $f$-flux. When one performs T-duality transformations on this space along the directions $x^2$ and $x^3$, it is mapped to itself. T-duality along $x^{2}$ transfers it into: 
  \item {\it Single elliptic spaces} with $f=0$, $H\neq0$: Here the first and the third line in table~\ref{tab:tdualitychain} correspond to the same non-geometric space with $H$- and $Q$-flux. The second line is the geometric background with $f$-Flux only, whereas the forth line corresponds to a non-geometric space with $f$- and with $R$-flux.
  \item {\it Double elliptic spaces} with $f\neq0$, $H\neq0$: 
Now all configurations in this table have a geometric and a non-geometric flux turned on at the same time. Here there is no T-dual configuration with geometric fluxes only. Hence the double elliptic spaces cannot be handled with standard supergravity; they always need a full DFT description.
\end{itemize}
The most interesting background is the double elliptic space, because it can not be described by SUGRA. Nevertheless, it is known from CFT \cite{Dabholkar:2002sy,Condeescu:2012sp,Condeescu:2013yma} and was discussed recently by \cite{ReviewDFT:2013hlb} in the context of large generalized diffeomorphisms in DFT.

\subsection{Background fields and field redefinitions}
In this final section of the paper we want to derive explicit expressions for the background fields, namely the metric, the $B$-field and the $\beta$-field, as functions of the doubled coordinates $Y^N$. We will focus on the double elliptic background, discussed in the last chapter. The fields of this background depend on one single coordinate direction, $y^1$ (or in a T-dual frame $\tilde y^1$), only. As usual, the expressions for the background fields are subject to possible field redefinitions, as used in \cite{Andriot:2011uh, Andriot:2012wx, Andriot:2012an}. These field redefinitions for example exchange the $B$-field with the $\beta$-field or vice versa. In this context it is a crucial question whether there is a certain field redefinition after which the background is a geometric space. As we will discuss, this is impossible for the double elliptic background, which is not T-dual to a geometric space.

As explained in section~\ref{sec:dftandsym}, the generalized vielbein $E\indices{^A_M}$ of the fiber is subject to a local double Lorentz symmetry, connecting
\begin{equation}
  \tilde{E}\indices{^A_M} = T\indices{^A_B} \widehat{E}\indices{^B_N} U\indices{^N_M}
    \quad \text{and} \quad
  E\indices{^A_M} = \widehat{E}\indices{^A_N} U\indices{^N_M}\;. 
\end{equation}
Here $T\indices{^A_B}$ is a double Lorentz transformation of the fiber, parameterized by $d_\mathrm{f}(d_\mathrm{f}-1)$ independent variables. All frames related via such transformations are physically equivalent. The twist \eqref{eqn:twistHf}, which was obtained in the last section, is an element of the double Lorentz group, too. For the vacuum, where $\widehat{E}\indices{^A_M}=\widehat{\bar E}\indices{^A_M}=\delta^A_M$, we are able to choose $T\indices{^A_B}$ as the inverse of the twist. In this case the generalized vielbein describes locally a flat space without fluxes. At first glance, this result seems strange. Because, we started explicitly with non-vanishing covariant fluxes in order to obtain a non-abelian gauge symmetry in the effective theory. This ambiguity is resolved when remembering that the background has a global monodromy, which can not be removed by local transformations on a single patch. A background which exhibits exactly this monodromy is the orbifold
\begin{equation}
  T^{4} / \mathrm{Z}_R \times \mathrm{Z}_L
    \quad\text{with}\quad
  R = \frac{1}{(f-H)\bmod 1}
    \quad\text{and}\quad
  L = \frac{1}{(f+H)\bmod 1}\,,
\end{equation}
where $H$ and $f$ are the fluxes we started with. The first discrete group acts on the right movers and the second one on the left movers. A setup with vanishing $f$ component, is a completely asymmetric orbifold, while a vanishing $H$ component leads to a symmetric orbifold. Locally, we are not able to distinguish it from a flat torus. Both are Ricci flat and satisfy the field equations. Nevertheless, globally they are very different. This observation emphasizes that the fluxes we started with play a significant r\^ole and are not only an unphysical gauge.

Before reading off the fields $\beta^{\mu\nu}$, $B_{\mu\nu}$ and the metric $g_{\mu\nu}$ from the generalized vielbein $E\indices{^A_M}$ in its most general parameterization \eqref{eqn:EAMgeneral}, we will fix the local double Lorentz symmetry. In general, there are two different possibilities to do so. The first and simplest one is the trivial choice $T\indices{^A_B}=\delta^A_B$. In this case one gets 
\begin{gather}
  B_{23} = -B_{32} = - \tan H x^1 \,, \quad \beta_{23} = - \beta_{32} = \frac{1}{2} \sin 2 H x^1 \quad \text{and} \quad \nonumber \\
  e\indices{^a_i} = \begin{pmatrix}
    1 & 0 & 0 \\
    0 & \frac{\cos f x^1}{\cos H x^1}  & \frac{\sin f x^1}{\cos H x^1} \\
    0 & -\frac{\cos f x^1}{\cos H x^1} & \frac{\cos f x^1}{\cos H x^1}
  \end{pmatrix}\,.
\end{gather}
For more sophisticated double Lorentz gauge fixings, we have to choose a different $T\indices{_A^B}$ at each point of the base. This choice should be done in such a way that it leaves some functions of components of the generalized vielbein $\tilde E\indices{^A_M}$, like e.g. $f_i(\tilde E\indices{^A_M})$, constant over the whole base. Technically speaking, this means $\partial_{\tilde L} f_i (\tilde E\indices{^A_M})=0$ has to vanish for all directions $\tilde L$ along the base. To evaluate such conditions, we start by calculating
\begin{equation}\label{eqn:localvarEAM}
  \partial_{\tilde L} \tilde{E}\indices{^A_M} =
    \partial_{\tilde L} T\indices{^A_B} \widehat{E}\indices{^B_N} U\indices{^N_M} +
    T\indices{^A_B} \widehat{E}\indices{^B_N} \partial_{\tilde L} U\indices{^N_M}\,.
\end{equation}
Furthermore we parameterize $T\indices{^A_M}$ in a similar way, as we have done it for $U\indices{^M_N}$ in \eqref{eqn:twistfromflux}. This gives rise to
\begin{equation}
  T\indices{^A_B} = \exp\left[ \mathcal{G}\indices{^A_B} (X^{\tilde L})  \right]
\end{equation}
where the arbitrary functions $\mathcal{G}\indices{^A_B}(X^{\tilde L})$ in bared indices have to fulfill
\begin{equation}
  \mathcal{G}\indices{^{\bar a}^{\bar b}} =
    \mathcal{G}\indices{_{\bar a}_{\bar b}} = 0
\end{equation}
in order to restrict $T\indices{^A_B}$ to the double Lorentz subgroup of the full O$(d_\mathrm{f}, d_\mathrm{f})$. The most significant difference between this definition and \eqref{eqn:twistfromflux} is that the exponent here is not restricted to a linear dependence in the base coordinates $X^{\tilde L}$. With this definitions at hand, \eqref{eqn:localvarEAM} equals
\begin{equation}\label{eqn:DGLfieldredef}
  \partial_{\tilde L} E\indices{^A_M} = T\indices{^A_B} \left( \partial_{\tilde L} 
    \mathcal{G}\indices{^B_C} \widehat{E}\indices{^C_N} + 
    \widehat{E}\indices{^B_K} \mathcal{F}\indices{^K_N_{\tilde L}} \right)
    U\indices{^N_M} \,.
\end{equation}
Let us now define the constituents of the generalized vielbein in \eqref{eqn:EAMgeneral} as
\begin{equation}
  e\indices{^a_i} = \begin{pmatrix}
    e\indices{^1_1} & e\indices{^1_2} \\ 0 & e\indices{^2_2}
  \end{pmatrix}\,, \quad
    B_{ij} = \begin{pmatrix}
    0 & B \\ -B & 0
  \end{pmatrix} \quad \text{and} \quad
  \beta^{ij} = \begin{pmatrix}
    0 & \beta \\ -\beta & 0
  \end{pmatrix}
\end{equation}
for our $d_\mathrm{f}=2$ example from the last section. This gives rise to
\begin{equation}
  e\indices{^1_1} = \frac{1}{E\indices{_1^1}}\,, \quad
  e\indices{^1_2} = - \frac{E\indices{_2^1}}{E\indices{_1^1} E\indices{_2^2}}\,, \quad
  e\indices{^2_2} = \frac{1}{E\indices{_2^2}}\,, \quad
  B = \frac{E_{12}}{E\indices{_1^1}} \quad \text{and} \quad
  \beta = E^{12} E\indices{_1^1}\,.
\end{equation}
In the following we use the three different derivatives:
\begin{align}\label{eqn:varB}
  \partial_1 B &= \frac{1}{E\indices{_1^1}} \left( \partial_1 E_{12} - \frac{E_{12}}{E\indices{_1^1}} 
    \partial_1 E\indices{_1^1}\right) \,, \\ \label{eqn:varbeta}
  \partial_1 \beta &= E\indices{_1^1} \partial_1 E^{12} + E^{12} \partial_1 E\indices{_1^1} 
    \quad \text{and} \\ \label{eqn:varVolume}
  \partial_1 \det ( e\indices{^a_i} ) &= - \partial_1 \frac{1}{E\indices{_1^1}E\indices{_2^2}}
    = \frac{1}{\left( E\indices{_1^1} E\indices{_2^2} \right)^2} \left(
    E\indices{_2^2} \partial_1 E\indices{_1^1} + E\indices{_1^1} \partial_1 E\indices{_2^2} \right)\,.
\end{align}
Setting one of them to zero, and using the derivative of the generalized vielbein \eqref{eqn:localvarEAM} gives rise to a differential equation for $\mathcal{G}\indices{^A_B}(y^1)$, parameterized by
\begin{equation}
  \mathcal{G}\indices{^A_B} = \frac{1}{2} \begin{pmatrix}
    0 & \xi(y^1) + \phi(y^1) & 0 & -\xi(y^1) + \phi(y^1) \\
    -\xi(y^1) -\phi(y^1) & 0 & \xi(y^1) - \phi(y^1) & 0 \\
    0 & -\xi(y^1) + \phi(y^1) & 0 & \xi(y^1) + \phi(y^1) \\
    \xi(y^1) - \phi(y^1) & 0 & -\xi(y^1) + \phi(y^1) & 0
  \end{pmatrix}\,.
\end{equation}
To obtain both parameters of the double Lorentz transformation, $\xi(x^1)$ and $\phi(x^1)$, one differential equation is not enough. Hence, we set additionally the derivative
\begin{equation}
  \partial_1 E\indices{^2_1} = 0
\end{equation}
to zero. This restricts the vielbein $e\indices{^a_i}$ to an upper triangular matrix and leads to a complete set of two coupled ordinary differential equations for $\xi$ and $\phi$. They can be solved numerically and depending on which of the derivatives \eqref{eqn:varB} - \eqref{eqn:varVolume} is set to zero, one obtains a totally double Lorentz fixed generalized vielbein $\tilde E\indices{^A_M}$ with
\begin{itemize}
  \item with constant $B$ (which we choose $B=0$)\,,
  \item with constant $\beta$ (which we choose $\beta=0$) or 
  \item with constant volume $V=\det ( e\indices{^a_i} )$ of the fiber.
\end{itemize}
These three choices are connected to each other via field redefinitions. For all $\widehat{E}\indices{^A_M} \ne \widehat{\bar E}\indices{^A_M}$, the first two cases lead to a metric with a discontinuity after one complete cycle around the base. Thus the field configurations obtained in this way, do not permit a geometric description and therefore are called non-geometric. Nevertheless, the question arises, whether there exists a field redefinition leading to a geometric description. This question naturally arises, because recent works like \cite{Andriot:2011uh,Andriot:2012wx} showed that certain backgrounds are non-geometric for the $\beta=0$ choice, but become geometric for $B=0$.

In order to find a field redefinition which leads to a geometric setup, one first has to formulate a criterion to distinguish between geometric and non-geometric configurations: For a geometric configuration, the monodromy
of the vielbein $e\indices{^a_i}$ has to be an element of the group of large diffeomorphisms on the torus. For $d_\mathrm{f}=2$, this group is SL$(2,Z)$ and one obtains the condition
\begin{equation}
  M\indices{^i_j} = e\indices{_a^i}(y^1)  e\indices{^a_j}(y^1 + 2\pi) \in \mathrm{SL}(2,Z)\,.
\end{equation}
It can only hold, if 
\begin{equation}\label{eqn:detofmonodromy}
  \det ( M\indices{^i_j} ) = \frac{V(2 \pi + y^1)}{V(y^1)} = 1 \quad
    \leftrightarrow \quad
  V(2 \pi + y^1) = V(y^1)
\end{equation}
is fulfilled. But for $B=0$ or $\beta=0$ this condition is violated. Thus the metric becomes discontinuous and prohibits a geometric description. This observation justifies the third case $V=$constant for which \eqref{eqn:detofmonodromy} is trivially fulfilled. With this fixing, which is implemented by setting $e\indices{^2_2}=V / e\indices{^1_1}$, the monodromy $M\indices{^i_j}$ reads
\begin{equation}\label{eqn:monodromyV=const}
  M\indices{^i_j} = \begin{pmatrix}
    \frac{e\indices{^1_1}(2\pi + y^1)}{e\indices{^1_1}(y^1)} & 
    \frac{e\indices{^1_2}(y^1 + 2\pi)}{e\indices{^1_1}(y^1)} 
    - \frac{e\indices{^1_2}(y^1)}{e\indices{^1_1}(y^1 + 2\pi)} \\
    0 & \frac{e\indices{^1_1}(y^1)}{e\indices{^1_1}(2\pi + y^1)}
  \end{pmatrix}\,.
\end{equation}

The differential equation, discussed above, is a straightforward approach to fix the double Lorentz symmetry, but it is not well suited for more general calculations. Thus we want to discuss another technique, which leads to the same results. It is based on the complex structure $\tau=\tau_\compliR + i \tau_\compliI$ and the K\"ahler parameter $\rho=\rho_\compliR + i \rho_\compliI$ of the fiber torus. By using the decomposition \eqref{eqn:decompSL2xSL2} we find
\begin{equation}
  E_\rho = \pm \frac{1}{\sqrt{V}}\begin{pmatrix}
    1 & B \\ - V \beta & 1 - B \beta
  \end{pmatrix} \quad \text{and} \quad
  E_\tau = \pm \frac{1}{\sqrt{V}}\begin{pmatrix}
    \frac{V}{e\indices{^1_1}} & 0 \\
    e\indices{^1_2} & e\indices{^1_1}
  \end{pmatrix}\,.
\end{equation}
With \eqref{eqn:actiontau&rho}, we can assign
\begin{align}
  \rho &= \frac{1}{\sqrt{V}} \cdot \frac{ \pm i + B }{
    \mp V \beta i + 1 - B \beta} &\text{and}& &
  \tau &= \frac{1}{\sqrt{V}} \cdot \frac{ \frac{V}{
    e\indices{^1_1}} i }{\pm e\indices{^1_2} i + e\indices{^1_1}}\,,
\intertext{Solving these two equations for $B$, $\beta$, $e\indices{^1_1}$ and $e\indices{^1_2}$ gives rise to}
  B &= \pm \frac{\sqrt{\pm \rho_\compliI \left( \rho_\compliI  \mp 
    V \left|\rho\right|^2 \right)}}{\rho_\compliI} \,, && &
  \beta &= \pm \frac{- \rho_\compliR + \sqrt{\pm \rho_\compliI \left( 
    \rho_\compliI  \mp  V \left|\rho\right|^2 \right)}}{
    V \left| \rho \right|^2} \\
  e\indices{^1_1} &= \pm \sqrt{\frac{\pm V \tau_\compliI}{\left| \tau \right|^2}}
    &\text{and}& & 
  e\indices{^1_2} &= \mp \sqrt{\frac{V}{\pm \tau_\compliI
    \left| \tau \right|^2}} \tau_\compliR \,.
\end{align}
The vielbein components $e\indices{^1_1}$ and $e\indices{^1_2}$ are defined for all $\tau \in \mathds{C}$. For $B$ and $\beta$, this is not the case. They are only defined in the complex region
\begin{equation}
  \mathds{C} \, \setminus \, \left\{ \rho \in \mathds{C} \, | \,
  \left |\rho - \frac{i}{2 V} \right| < \frac{1}{2 V} \lor 
  \left |\rho + \frac{i}{2 V} \right| < \frac{1}{2 V} \right\}\,.
\end{equation}
In order show the implications of this constraint, we consider a $\rho(0)=\exp(i \theta)$ where $0 \le \theta \le \frac{\pi}{2}$. From \eqref{eqn:Utau&rho} it follows that the complex function $\rho(y^1)$ is given by
\begin{equation}
  \rho(y^1) = \frac{ \rho(0) \cos(H y^1) + \sin(H y^1) }{
    - \rho(0) \sin(H y^1) + \cos(H y^1)}\,.
\end{equation}
In the complex plane, all possible values of this function lay on a circle around the point
\begin{equation}
  z = i z_\compliI = \frac{i \rho_\compliI(0)}{1 - \rho_\compliR^2(0)} = \frac{i}{\sin \theta} \quad
    \text{which has the radius} \quad
    R = \frac{ \rho_\compliI(0) \, \rho_\compliR(0)}{1 - \rho_\compliR^2(0)} = \cot \theta \,.
\end{equation}
Because $z_\compliI > R$ we only need to consider the upper half of the complex plane. The circle with center $z$ and radius $R$ must not intersect the region where $B$ and $\beta$ are not defined. Thus one has to constrain the volume $V$ of the fiber to
\begin{equation}
  V \le z_\mathrm{I} - R = \frac{\rho_\compliI(0)}{1 - \rho_\compliR(0)}=\frac{\sin\theta}{1-\cos\theta}\,.
\end{equation}
This fact is important, because it shows that when fixing the volume $V$ of the fiber to a finite value, there are always some field configurations which are not well defined in terms of $B$ and $\beta$.

Finally we discuss the monodromy $M\indices{^i_j}$ for $f=1/4$. In this case, the twist gives rise to
\begin{equation}
  \tau(2\pi + y^1) = - \frac{1}{\tau(y^1)}\,.
\end{equation}
Taking into account that
\begin{equation}
  e\indices{^a_i} = \sqrt{\frac{V}{\tau_\compliI}} \begin{pmatrix}
    \tau_\compliI & \tau_\compliR \\
    0 & 1
  \end{pmatrix}\,,
\end{equation}
one obtains
\begin{equation}
  e\indices{^1_1}(2\pi) = \frac{1}{|\tau(0)|} e\indices{^1_1}(0) \quad \text{and} \quad
  e\indices{^1_2}(2\pi) = - \frac{1}{|\tau(0)|} e\indices{^1_2}(0)\,.
\end{equation}
Plugging this result into \eqref{eqn:monodromyV=const} gives rise to
\begin{equation}
  M\indices{^i_j} = \frac{1}{\left| \tau(0) \right|} \begin{pmatrix}
    1 & - \frac{\tau_\compliR(0)}{\tau_\compliI(0)}\left(|\tau(0)|^2 + 
      1 \right) \\  0 & \left| \tau(0) \right|^2  
  \end{pmatrix}\,.
\end{equation}
Now there are two possibilities: $M\indices{^i_j}$ itself is an SL$(2,Z)$ matrix, or it can be transformed by a GL$(d)$ transformation $t\indices{^i_j}$ into such a matrix. GL$(d)$ transformations act as
\begin{equation}
  \tilde{M}\indices{^i_j} = t\indices{^i_k} M\indices{^k_l} t\indices{_j^l}
\end{equation}
on the monodromy. In accordance with the notation uses so far, $t\indices{_i^j}$ is the inverse, transpose of $t\indices{^i_j}$. Such a transformation only exists, when the trace of $M$ is an integer, namely
\begin{equation}
  \Tr M\indices{^i_j} = \left( | \tau(0) | + \frac{1}{| \tau(0) |} \right) \in \mathds{Z}\,.
\end{equation}
There are some special points for which this constraint hold, but in general it is violated and one ends with a non-geometric background as expected.

\section{Conclusions and discussion}\label{sec:conclusions}
In this paper we have applied a consistent Scherk-Schwarz ansatz to Double Field Theory in order to construct 
a reduced effective theory. This effective theory is used to find
\begin{enumerate}
  \item non-trivial vacuum solutions of DFT's equations of motion and
  \item to describes fluctuations around this vacuum.
\end{enumerate}
To do this, we use a generalization of group manifolds, which are well understood for ordinary geometry, but has to be adapted to DFT. These manifolds need to have as many isometries as coordinates. In DFT, isometries are defined by the vanishing generalized Lie derivatives, 
\begin{equation}
  \mathcal{L}_{K\indices{_I^J}} \mathcal{H}^{MN} = 0 \quad \text{and} \quad \mathcal{L}_{K\indices{_I^J}} \phi' = 0\,.
\end{equation}
They give rise to homogeneous, doubled spaces which exhibit a constant generalized Ricci scalar (which is equivalent to the scalar potential in the effective theory). From the effective theory's point of view, these spaces are completely specified by the structure coefficients of the group they are linked to. The structure coefficients can be expressed in terms of the covariant fluxes $\mathcal{F}_{ABC}$. They are not arbitrary, but have to fulfill several constraints. In general, these constraints can be divided into three different categories: The first kind of constraints is needed to create a group structure. It requires that the covariant fluxes are constant and the Jacobi identity (or, more generally, the quadratic constraint) is fulfilled. Additionally, the second kind of constraints requires 
that the group manifold is compatible with the strong constraint. Such constraints are challenging, because the strong constraint has to be checked on the level of the generalized metric. But the map between covariant fluxes and generalized metric is involved, so in general one can only find conditions for the fluxes which lead to a violation of the strong constraint. Nevertheless, they help to restrict the number of covariant fluxes which survived the constraints of the first kind. Finally the field equations of the effective theory limit the allowed covariant fluxes. In this paper we looked  for a vacuum solution which gives rise to a Minkowski space in the external direction. Thus the scalar potential $V$ has to have a minimum with $V=0$. This again puts severe restrictions on the covariant fluxes.

In $D-d=3$, the only covariant fluxes which fulfill all constraints, discussed above, are
\begin{equation}
  H_{123} = Q^{23}_1 = H \quad  Q^{31}_2 = Q^{12}_3 = 0\,, \quad
  R^{123} = f^1_{23} = 0  \quad \text{and} \quad f^2_{31} = f^3_{12} = f \,.
\end{equation}
For them, we construct the twist $U\indices{^M_N}$ and the Killing vectors $K\indices{_I^J}$. Especially the Killing vectors are essential for a consistent dimensional reduction. In the literature they have not been discussed before. For $H\ne 0$ and $f\ne 0$, the background which corresponds to the fluxes above is not T-dual to a background with geometric fluxes only. In this case, the Killing vectors depend on the coordinates and the dual coordinates. They violate the strong constraint, but nevertheless the algebra generated by them is closed. These Killing vectors describe all three possible kinds of generalized diffeomorphism (coordinate transformations, $B$- and $\beta$-field gauge transformations) at the same time. Thus it is impossible to describe such background in SUGRA or generalized geometry. We also showed that it is impossible to find a field redefinition which makes the background and fluctuations around it well defined. Thus we come to the conclusion that these backgrounds are beyond the scope of SUGRA and generalized geometry.

We also considered fluctuations around these backgrounds which have the same isometries (Killing vectors) as the background itself. In terms of the effective actions such fluctuations can be expressed as $(D-d)^2$ scalar, and $2(D-d)$ vector bosons. For these bosons we calculated the mass spectrum and the gauge group. So we use DFT in a {\it twofold way}. First we use it to calculate the background and afterwards, it is used to study fluctuations around this background. This is possible because DFT is a background independent theory. So it not only makes predictions about valid backgrounds, but also about fluctuations around these background. The gaugings we found are compatible with the CFT description of asymmetric orbifold discussed in \cite{Condeescu:2013yma}. Furthermore, the way the twist $U\indices{^M_N}$ acts on the generalized vielbein suggests that the double elliptic background has a realization as an asymmetric orbifold in string theory.

Explicit CFT computations in this kind on string background could also confirm the mass spectrum we have calculated. This would be an important check that DFT indeed covers such string backgrounds. 

\acknowledgments
We gratefully acknowledge that Olaf Hohm was involved in the initial stages of this project. We like to thank him for many important discussions  during the preparation of this paper. We also would like to thank D. Andriot, A. Betz, R. Blumenhagen, S. Massai, F. Montiel, S. Nibbelink, P. Patalong and M. Schmidt Sommerfeld for helpful discussions. This work was partially supported by the ERC Advanced Grant ``Strings and Gravity''(Grant.No. 32004) and by the DFG cluster of excellence ``Origin and Structure of the Universe''.

\bibliography{literature}
\bibliographystyle{JHEP}

\end{document}